\documentclass[%
 aps,
 prb,
 amsmath,
 amssymb,
 showpacs,
 twocolumn
]{revtex4-1}

\usepackage{graphicx}
\usepackage{dcolumn}
\usepackage{bm}
\usepackage{color}

\begin{document}


\title{Spontaneous emission from large quantum dots in nanostructures: exciton-photon interaction beyond the dipole approximation}

\author{S. Stobbe$^1$}\email[]{stobbe@nbi.ku.dk}
\author{P. T. Kristensen$^2$}
\author{J. E. Mortensen$^2$}
\author{J. M. Hvam$^2$}
\author{J. M{\o}rk$^2$}
\author{P. Lodahl$^1$}
\affiliation{
$^{1}$Niels Bohr Institute, University of Copenhagen, Blegdamsvej 17, DK-2100 Copenhagen, Denmark\\
$^{2}$DTU Fotonik, Department of Photonics Engineering, Technical University of Denmark, {\O}rsteds Plads 343, DK-2800 Kgs.~Lyngby, Denmark
}

\date{\today}

\begin{abstract}
We derive a rigorous theory of the interaction between photons and spatially extended excitons confined in quantum dots in inhomogeneous photonic materials. We show that, beyond the dipole approximation, the radiative decay rate is proportional to a non-local interaction function, which describes the interaction between light and spatially extended excitons. In this regime, light and matter degrees of freedom cannot be separated and a complex interplay between the nanostructured optical environment and the exciton envelope function emerges. We illustrate this by specific examples and derive a series of important analytical relations, which are useful for applying the formalism to practical problems. In the dipole limit, the decay rate is proportional to the projected local density of optical states and we obtain the strong and weak confinement regimes as special cases.
\end{abstract}


\maketitle

\section{Introduction}
The dipole approximation (DA) is one of the most central and successful approximations in quantum optics and quantum electrodynamics (QED). When describing the light-matter interaction, the DA is valid if the variation of the electromagnetic field is negligible over the spatial extent of the emitter. Since optical wavelengths exceed atomic dimensions by orders of magnitude, this is an excellent approximation in atomic physics. The advances in solid-state quantum optics have enabled the realization of semiconductor nanostructures with strongly modified optical properties and embedded self-assembled quantum dots (QDs). Both atoms and QDs have a discrete spectrum with optically active transitions but as opposed to atoms, QDs are inherently mesoscopic solid-state structures whose transition energy, position, and chemical composition can be controlled by semiconductor nanotechnology, and in nanophotonic structures the electromagnetic environment can have pronounced spatial variations. For QDs the validity of the DA is not clear \emph{a priori}, and the purpose of this article is to derive the theory of spontaneous emission beyond the DA for excitons confined in QDs embedded in nanostructures. Recently, a large deviation from dipole theory was observed for small QDs in close proximity to a metallic mirror~\cite{Andersen2011}, directly illustrating the need for a theory of light-matter interaction beyond the DA.

The tunability of QD sizes can lead to very interesting exciton effects. For strongly confined states in small QDs the Coulomb interaction can be neglected and excitons can thus be described as mutually independent electrons and holes~\cite{Efros1982,Schmitt-Rink1987}. For larger QDs the Coulomb interaction plays an increasingly important role and the electron and the hole form a bound exciton state, which has an oscillator strength (OS) proportional to the volume of the exciton. This is the so-called giant-OS effect~\cite{Elliott1957,Rashba1962,Hanamura1988}, which has received particular attention in the field of solid state QED because of the theoretical prediction~\cite{Andreani1999} that QDs must be in the giant-OS regime in order to achieve strong coupling between a single QD and a microcavity. Indeed, some of the first demonstrations of strong coupling in microcavities were achieved with large QDs~\cite{Reithmaier2004,Peter2005}. However, the rapid increase in the quality factors and the reduction of mode volumes in photonic crystal membrane nanocavities over the past years have enabled strong coupling using small QDs~\cite{Yoshie2004,Englund2007,Laucht2009}.

A key signature of the giant-OS effect is fast radiative decay rates~\cite{Stobbe2010}. Fast total decay rates have been observed~\cite{Hours2005,Reitzenstein2009} but the non-radiative decay rate was not measured in any studies of large QDs except for a recent work~\cite{Stobbe2010}. There it was shown that contrary to the common assumption, non-radiative recombination can be the dominant decay process for large QDs resulting in a small OS but a fast total decay rate. Therefore, measuring the non-radiative decay rate is essential to drawing conclusions about the OS and this has so far hindered a complete experimental demonstration of the giant-OS effect.

\begin{figure}[t!]
\begin{center}
\includegraphics[width=0.46\columnwidth]{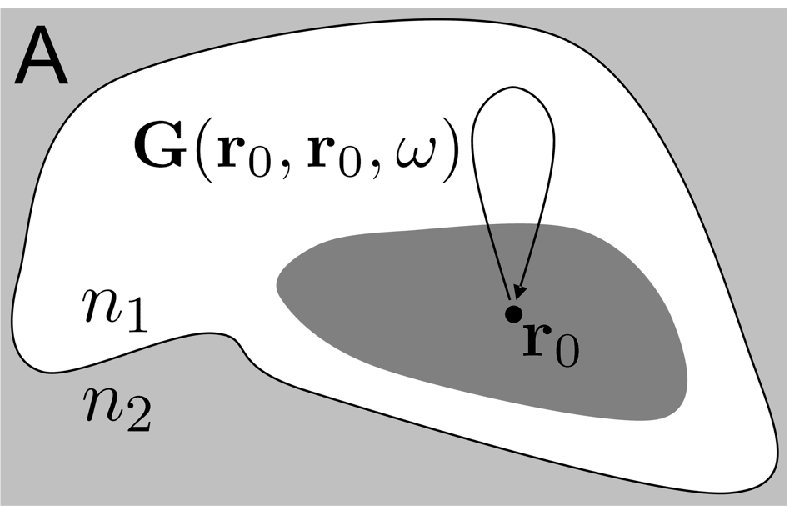}\hfill
\includegraphics[width=0.46\columnwidth]{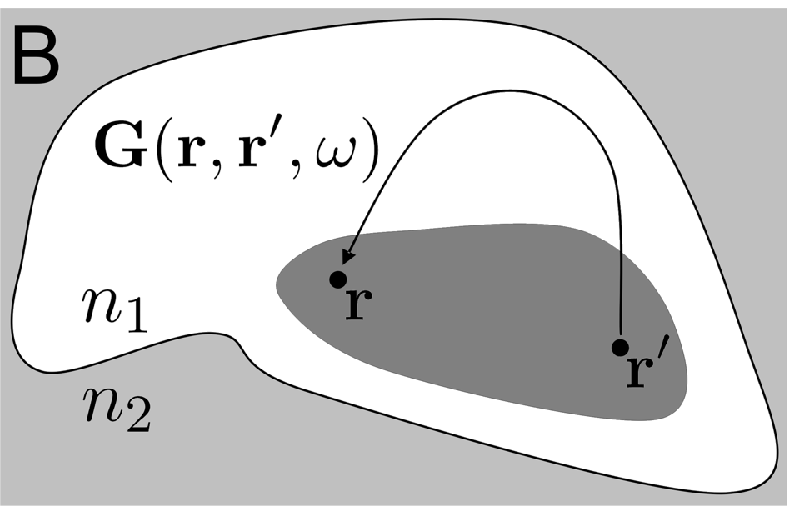}
\caption{Comparison between the LDOS and the non-local interaction function introduced in this work. An exciton emitter with a given envelope function (dark gray) is embedded in an inhomogeneous dielectric environment indicated schematically by the different refractive indices $n_1$ (white) and $n_2$ (light gray). The Green's tensor $\mathbf{G}(\mathbf{r},\mathbf{r}',\omega)$ is a propagator of the electric field between two spatial points at a given frequency $\omega$ and it is depicted as the arrows. (A) In the DA the spontaneous emission rate is governed by the LDOS, which is given by the imaginary part of the Green's tensor evaluated at $(\mathbf{r}_0,\mathbf{r}_0)$, where $\mathbf{r}_0$ is the center of the emitter. (B) Beyond the DA the spontaneous emission rate is governed by a non-local interaction function, which is given by an integral over the imaginary part of the Green's tensor connecting all possible combinations of $\mathbf{r}$ and $\mathbf{r}'$, weighted by the envelope function.
\label{fig:ComparisonBetweenLWAandBeyondLWA}}
\end{center}
\end{figure}

Another effect of increasing the QD size, which has received much less attention~\cite{Sugawara1995,Ahn2003,Ishihara2004}, is the fact that for sufficently large QDs the DA may break down. The usual criterion for the validity of the DA is that the product of the length of the optical wave vector $\mathbf{k}$ and the spatial extent of the emitter $\mathcal{L}$ must be much smaller than unity~\cite{Scully_and_Zubairy}, i.e., $|\mathbf{k}| \mathcal{L} \ll 1$, but this criterion is insufficient to ensure the validity of the DA for QDs in nanostructures. For QDs, there are four reasons why the DA could break down. Firstly, the QDs are embedded in semiconductors with a high refractive index, e.g., $n\approx 3.4$ for GaAs, which increases $|\mathbf{k}|$. Secondly, QDs can be as large as $ \mathcal{L}=100\:\mathrm{nm}$ in lateral size~\cite{Reithmaier2004}. As an example, for $ \mathcal{L}=45\:\mathrm{nm}$ and a free-space wavelength of 970\,nm we obtain $|\mathbf{k}| \mathcal{L}\approx 1$. Thirdly, the criterion stated above is valid for homogeneous media. In nanostructures the optical field modes can have strong gradients rendering the DA invalid even if $|\mathbf{k}| \mathcal{L}\ll 1$ is fulfilled in bulk for light at the same frequency. Fourthly, the influence of the finite size of the QDs is enhanced by the asymmetric nature of exciton wave functions in QDs~\cite{Andersen2011}. Thus, a proper theory of spontaneous emission beyond the DA must be valid for arbitrary electromagnetic environments. Here we derive such a theory from first principles and show that the radiative decay rate depends on a non-local interaction function, whose physical interpretation is illustrated in Fig.~\ref{fig:ComparisonBetweenLWAandBeyondLWA}. In the DA (Fig.~\ref{fig:ComparisonBetweenLWAandBeyondLWA}(A)) the radiative decay rate is proportional to the projected local density of optical states (LDOS), which describes the field amplitude at the position of the emitter due to emission from a dipole source at the same position. Beyond the DA (Fig.~\ref{fig:ComparisonBetweenLWAandBeyondLWA}(B)), a double integral over all points in space must be performed, where the integrand is weighted by the envelope function of the exciton. The double integral describes the physics emerging beyond the DA.
These effects find a natural description within the framework of the electromagnetic Green's tensor. Thus, the self-interference giving rise to spontaneous emission is described mathematically by the imaginary part of the electromagnetic Green's tensor.

The non-local aspect of light-matter interaction beyond the DA implies that light and matter degrees of freedom cannot be separated, i.e., the radiative decay rate is neither proportional to the projected LDOS nor to the OS. This points to another previously overlooked problem in the interpretation of the experimental results on large QDs: even if the non-radiative decay rate discussed above had been measured and found negligible, the highly non-trivial influence of the ubiquitous surrounding optical nanostructure cannot be approximated by a homogeneous medium and the OS has no general physical meaning. Here we present the complete quantum theory of spontaneous emission for two-level QDs in inhomogeneous media, which provides the theoretical framework for more quantitative future experiments and enables calculating non-Markovian decay dynamics and radiative (Lamb) shifts. We consider InGaAs QDs, but our formalism can be readily modified to describe other materials.

This paper is organized as follows: In section I we describe the exciton state and in section II we calculate the spontaneous emission from excitons beyond the DA in the Wigner-Weisskopf model. In section III we derive the connection to dyadic Green's tensors, introduce a non-local interaction function and discuss the physical implications of the results. We consider the dipole limit of our results in section IV and we apply our formalism to three special cases in section V. Finally, we present the conclusions in section VI. In Appendix A we solve the effective mass equation for the geometries relevant for this work. In Appendix B we consider the classical analogue of our results. The derivation of the relation to Green's tensors is included in Appendix C. Finally, in Appendix D we show the analytical calculation of the decay rate of spherical excitons beyond the DA.

\section{Excitons in quantum dots}
A bulk semiconductor consists of nuclei and electrons and in the Born-Oppenheimer approximation the motion of nuclei and electrons are decoupled. Thus we consider electron states imposed on an equilibrium state of the nuclei. At low temperatures the valence bands are completely filled while the conduction bands are empty. Although the upper valence bands in InAs and GaAs are degenerate in bulk materials, this degeneracy is lifted in the presence of confinement and strain in QDs and we consider the heavy-hole band only. This is a good approximation for low excitation powers and low temperatures~\cite{Stobbe2009}.

The description of the ground state of a bulk semiconductor consisting of $N$ electrons in the upper valence band must be treated in a many-body formalism~\cite{Hanamura1988,Dimmock1967,Knox,Schmitt-Rink1989}. In the simplest possible case, i.e.,\ when neglecting interactions, the ground state wave function is given by the Slater determinant~\cite{Liboff}. We neglect the spin degree of freedom, which amounts to considering only bright excitons in which the electron and hole spins are antiparallel\cite{Poem2010,Johansen2010}. It is convenient to use the compact occupation number formalism and we define the ground state of the crystal as the Fermi sea $|\mathcal{F}\rangle$ given by
\begin{eqnarray}
|\mathcal{F} \rangle &=& |1_{\mathrm{v},\mathbf{k}_1},\ldots, 1_{\mathrm{v},\mathbf{k}_i}, \ldots , 1_{\mathrm{v},\mathbf{k}_N}\rangle\\
&=& \prod_{\mathbf{k}_i} c^\dagger_{\mathrm{v}, \mathbf{k}_i} | 0_\mathcal{F} \rangle,\nonumber
\end{eqnarray}
where $c_{\mathrm{v}, \mathbf{k}_i}$ ($c^\dagger_{\mathrm{v}, \mathbf{k}_i}$) is the annihilation (creation) operator of an electron in the valence band with $\mathbf{k} = \mathbf{k}_i$ and $| 0_\mathcal{F} \rangle$ denotes the state void of any electrons.
These second-quantization operators create single-particle states with corresponding wave functions $\psi_{\mathrm{v}, \mathbf{k}_i}(\mathbf{r}_i)$, which may be written in Bloch-form as
\begin{equation}
\psi_{\mathrm{v}, \mathbf{k}_i}(\mathbf{r}_i) = \frac{1}{\sqrt{V}} e^{\mathrm{i}\mathbf{k}_i\cdot\mathbf{r}_i}u_{\mathrm{v}, \mathbf{k}_i}(\mathbf{r}_i).\label{eq:Bloch}
\end{equation}
Here the valence band Bloch function $u_{\mathrm{v},\mathbf{k}_i}(\mathbf{r}_i)$ has the periodicity of the crystal lattice and is normalized over a unit cell and $V$ denotes the crystal volume.

We can write an excited state of the bulk semiconductor as
\begin{eqnarray}
| X_{\mathbf{k}_\mathrm{c}\mathbf{k}_\mathrm{v}} \rangle = c^\dagger_{\mathrm{c}, \mathbf{k}_\mathrm{c}} c_{\mathrm{v}, \mathbf{k}_\mathrm{v}}  | \mathcal{F} \rangle,\label{eq:Bulk_Exciton}
\end{eqnarray}
where $c_{\mathrm{c}, \mathbf{k}_\mathrm{c}}$ ($c^\dagger_{\mathrm{c}, \mathbf{k}_\mathrm{c}}$) is the annihilation (creation) operator of an electron in the conduction band with $\mathbf{k} = \mathbf{k}_\mathrm{c}$. The operators considered above are in the electron representation, but at this point it is convenient to change to the electron-hole representation by defining the following operators~\cite{Mattuck} $a_{\mathbf{k}_\mathrm{e}} = c_{\mathrm{c}, \mathbf{k}_\mathrm{c}}$, $a^\dagger_{\mathbf{k}_\mathrm{e}} = c^\dagger_{\mathrm{c}, \mathbf{k}_\mathrm{c}}$, $b_{\mathbf{k}_\mathrm{h}} = c^\dagger_{\mathrm{v}, \mathbf{k}_\mathrm{v}}$, and $b^\dagger_{\mathbf{k}_\mathrm{h}} = c_{\mathrm{v}, \mathbf{k}_\mathrm{v}}$, where the electron operators have simply been renamed and $b_{\mathbf{k}_\mathrm{h}}$ ($b^\dagger_{\mathbf{k}_\mathrm{h}}$) denotes the annihilation (creation) operator of a hole in the valence band. With this convention we write the excited state of the bulk semiconductor as
\begin{eqnarray}
| X_{\mathbf{k}_\mathrm{e}\mathbf{k}_\mathrm{h}} \rangle = a^\dagger_{\mathbf{k}_\mathrm{e}} b^\dagger_{\mathbf{k}_\mathrm{h}}  | \mathcal{F} \rangle.
\end{eqnarray}
This definition of the electron-hole representation has a number of consequences for the properties of holes. In particular, the following transformations hold, where the subscript $h$ refers to holes in the electron-hole representation and the subscript $v$ refers to electrons in the valence band in the electron representation~\cite{Mattuck,Kittel}: $\mathbf{k}_\mathrm{h} = -\mathbf{k}_\mathrm{v}$ (wave vector), $E_\mathrm{h,v} = - E_\mathrm{e,v}$ (energy), $m_\mathrm{h} = - m_\mathrm{v}$ (effective mass), $q_\mathrm{h} = -q$ (charge), and $V_\mathrm{h}(\mathbf{r}) = - V_\mathrm{v}(\mathbf{r})$   (confinement potential). Here $E_\mathrm{h,v}$ ($E_\mathrm{e,v}$) denotes the energy of a hole (an electron) relative to the valence band edge energy $E_\mathrm{v}$ and $q$ is the negative of the elementary charge, i.e., $q=-|e|$.

In the presence of Coulomb interaction and/or quantum confinement potentials the states $| X_{\mathbf{k}_\mathrm{e}\mathbf{k}_\mathrm{h}} \rangle$ are no longer eigenstates. Instead the new exciton eigenstate $| X \rangle$ can be expanded as
\begin{eqnarray}
| X \rangle = \sum_{\mathbf{k}_\mathrm{e},\mathbf{k}_\mathrm{h}} \tilde \chi_{\mathbf{k}_\mathrm{e},\mathbf{k}_\mathrm{h}} | X_{\mathbf{k}_\mathrm{e}\mathbf{k}_\mathrm{h}} \rangle,\label{eq:ExcitonExpansion}
\end{eqnarray}
where $\tilde \chi_{\mathbf{k}_\mathrm{e},\mathbf{k}_\mathrm{h}}$ are expansion coefficients. The corresponding wave function $X(\mathbf{r}_0,\mathbf{r}_\mathrm{e},\mathbf{r}_\mathrm{h})$ can be found by projection onto the position eigenvectors, where we have explicitly included the center position of the QD, $\mathbf{r}_0$,
\begin{align}
X(\mathbf{r}_0,\mathbf{r}_\mathrm{e},\mathbf{r}_\mathrm{h}) =& \sum_{\mathbf{k}_\mathrm{e},\mathbf{k}_\mathrm{h}} \tilde\chi_{\mathbf{k}_\mathrm{e},\mathbf{k}_\mathrm{h}}  \langle \mathbf{r}_\mathrm{e} \mathbf{r}_\mathrm{h} | X_{\mathbf{k}_\mathrm{e}\mathbf{k}_\mathrm{h}} \rangle \nonumber \\
=& \frac{1}{V}\sum_{\mathbf{k}_\mathrm{e},\mathbf{k}_\mathrm{h}} \tilde\chi_{\mathbf{k}_\mathrm{e},\mathbf{k}_\mathrm{h}} \nonumber\\
  & \times e^{\mathrm{i}\mathbf{k}_\mathrm{e}\cdot\mathbf{r}_\mathrm{e}} u_{\mathrm{c},\mathbf{k}_\mathrm{e}}(\mathbf{r}_\mathrm{e})  e^{\mathrm{i}\mathbf{k}_\mathrm{h}\cdot\mathbf{r}_\mathrm{h}} u_{\mathrm{v},\mathbf{k}_\mathrm{h}}(\mathbf{r}_\mathrm{h})\label{eq:ExcitonWavefunction}
\\
\simeq & \chi(\mathbf{r}_0,\mathbf{r}_\mathrm{e},\mathbf{r}_\mathrm{h}) u_{\mathrm{c},\mathbf{0}}(\mathbf{r}_\mathrm{e}) u_{\mathrm{v},\mathbf{0}}(\mathbf{r}_\mathrm{h}),
\end{align}
where Bloch's theorem, the transformation of a sum to an integral, $\sum_\mathbf{k}\to\frac{V}{(2\pi)^3}\int\mathrm{d}\mathbf{k}$, and the definition of an inverse Fourier transform have been used. The last equality holds when only excitations near the band edge are considered so that the Bloch functions may be evaluated at $\mathbf{k} = \mathbf{0}$, which is a good approximation for low temperatures and low excitation intensities. The function $\chi(\mathbf{r}_0,\mathbf{r}_\mathrm{e},\mathbf{r}_\mathrm{h})$ is denoted the exciton envelope function, which is given by the solution to the effective mass equation~\cite{Dimmock1967}
\begin{equation}
H_\mathrm{EM}(\mathbf{r}_0,\mathbf{r}_\mathrm{e},\mathbf{r}_\mathrm{h}) \chi(\mathbf{r}_0,\mathbf{r}_\mathrm{e},\mathbf{r}_\mathrm{h}) = (E-E_\mathrm{g}) \chi(\mathbf{r}_0,\mathbf{r}_\mathrm{e},\mathbf{r}_\mathrm{h}),\label{eq:EffectiveMassEquation}
\end{equation}
where $E$ is the exciton energy, $E_\mathrm{g} = E_\mathrm{c} - E_\mathrm{v}$ is the band gap energy, and the effective-mass Hamiltonian is given by
\begin{equation}
\begin{split}
H_\mathrm{EM}(\mathbf{r}_0,\mathbf{r}_\mathrm{e},\mathbf{r}_\mathrm{h}) =& \frac{\mathbf{p}_\mathrm{e}^2}{2m_0m_\mathrm{e}} + \frac{\mathbf{p}_\mathrm{h}^2}{2m_0m_\mathrm{h}} + V_\mathrm{e}(\mathbf{r}_0,\mathbf{r}_\mathrm{e})\\& + V_\mathrm{h}(\mathbf{r}_0,\mathbf{r}_\mathrm{h})
 - \frac{q^2}{4 \pi \epsilon_0 \epsilon_\mathrm{r} |\mathbf{r}_\mathrm{e} - \mathbf{r}_\mathrm{h}|}\label{eq:EffectiveMassHamiltonian}.
\end{split}
\end{equation}
Here $\mathbf{p}_\mathrm{e}$ ($\mathbf{p}_\mathrm{h}$) is the electron (hole) momentum operator, $m_0$ is the electron rest mass, $\epsilon_0$ is the vacuum permittivity, and $\epsilon_\mathrm{r}$ is the relative dielectric constant. In the strong-confinement limit where the Coulomb interaction can be neglected the solution to Eq.~ is $\chi(\mathbf{r}_0,\mathbf{r}_\mathrm{e},\mathbf{r}_\mathrm{h}) = F_\mathrm{e}(\mathbf{r}_0,\mathbf{r}_\mathrm{e}) F_\mathrm{h}(\mathbf{r}_0,\mathbf{r}_\mathrm{h})$, where $F_\mathrm{e}(\mathbf{r}_0,\mathbf{r}_\mathrm{e})$ and $F_\mathrm{h}(\mathbf{r}_0,\mathbf{r}_\mathrm{h})$ denote the electron and hole envelope functions, respectively. In the weak-confinement limit, electrons and holes are entangled and therefore their wave functions do not separate. We consider solutions to the effective-mass equation in specific geometries in Appendix~\ref{sec:confinement_regimes}.

\section{Quantum theory of spontaneous emission beyond the dipole approximation\label{sec:QuantumTheory}}
We describe light-matter interaction by the minimal coupling Hamiltonian in the generalized Coulomb gauge in which we assume $\nabla \cdot \left( \epsilon_\mathrm{r} (\mathbf{r}) \mathbf{A}(\mathbf{r},t) \right) = 0$, where $\mathbf{A}(\mathbf{r},t)$ is the vector potential. The interaction Hamiltonian reads~\cite{Cohen-Tannoudji,Vats2002}
\begin{align}
H'(\mathbf{r},t) =& \frac{\mathrm{i} \hbar q}{m_0} \mathbf{A}(\mathbf{r},t) \cdot \nabla.\label{eq:Hamltionian-p.A}
\end{align}
The vector potential is given by~\cite{Thirunamachandran}
\begin{equation}
\mathbf{A}(\mathbf{r},t)= \sum_{\mathbf{\mu}} \frac{\epsilon_\mathbf{\mu}}{\omega_\mathbf{\mu}} \hat{\mathbf{e}}_\mathbf{\mu} \left( A_\mathbf{\mu}(\mathbf{r}) a_\mathbf{\mu} e^{-\mathrm{i}\omega_\mathbf{\mu} t}+ A^\ast_\mathbf{\mu}(\mathbf{r}) a^\dagger_\mathbf{\mu} e^{\mathrm{i}\omega_\mathbf{\mu} t} \right), \label{eq:ApotentialQuantum1}
\end{equation}
where $\mathbf{\mu} = (\mathbf{k},s)$ is the combined wavevector $\mathbf{k}$ and polarization index $s \in \{ 1,2 \}$, $\omega_\mathbf{\mu}$ is the optical angular frequency, $\epsilon_\mathbf{\mu} = \sqrt{\frac{\hbar \omega_\mathbf{\mu}}{2 \epsilon_0}}$ is a normalization constant, $\hat{\mathbf{e}}_\mathbf{\mu}$ is the polarization unit vector, and $A_\mathbf{\mu}(\mathbf{r})$ is the field distribution function that solves the vector Helmholtz equation with fixed boundary conditions. $a_{\mathbf{\mu}}$ and $a_{\mathbf{\mu}}^\dagger$ are the field annihilation and creation operators, respectively.
In second quantization the interaction Hamiltonian can be written as
\begin{equation}
H' = \sum_{\mathbf{k}_i,\mathbf{k}_j} \sum_{\alpha,\beta} H'^{\alpha\beta}_{\mathbf{k}_i,\mathbf{k}_j} c^\dagger_{\alpha \mathbf{k}_i} c_{\beta \mathbf{k}_j},\label{eq:Hamiltonian_SecondQ}
\end{equation}
where $\alpha,\beta \in \{\mathrm{c},\mathrm{v}\}$ and
$H'^{\alpha\beta}_{\mathbf{k}_i,\mathbf{k}_j} = \langle 1_{\alpha,\mathbf{k}_i} | H'(\mathbf{r},t) | 1_{\beta,\mathbf{k}_j} \rangle$.

It is convenient to use the interaction picture where the time-evolution of the operators is governed by the non-interacting Hamiltonian. When considering only excitations near the band edge $\mathbf{k} = \mathbf{0}$, the energies of the conduction and valence bands and hence the transition energy of the QD, $\hbar \omega_0$, do not depend on $i$ or $j$. The operators in the interaction picture become $\tilde c_{\mathrm{c},\mathbf{k}} = c_{\mathrm{c},\mathbf{k}} e^{-\mathrm{i}\omega_\mathrm{c} t}$, $\tilde c^\dagger_{\mathrm{c},\mathbf{k}} = c^\dagger_{\mathrm{c},\mathbf{k}} e^{\mathrm{i}\omega_\mathrm{c} t}$, $\tilde c_{\mathrm{v},\mathbf{k}} = c_{\mathrm{v},\mathbf{k}} e^{-\mathrm{i}\omega_\mathrm{v} t}$, and $\tilde c^\dagger_{\mathrm{v},\mathbf{k}} = c^\dagger_{\mathrm{v},\mathbf{k}} e^{\mathrm{i}\omega_\mathrm{v} t}$. When inserting these operators in Eq.~(\ref{eq:Hamiltonian_SecondQ}) we obtain terms proportional to $e^{\pm \mathrm{i} (\omega_\mathbf{\mu} + \omega_0) t}$ and $e^{\pm \mathrm{i}\omega_\mathbf{\mu} t}$, which are rapidly oscillating as a function of time as well as the slowly oscillating terms proportional to $e^{\pm \mathrm{i} \Delta_\mathbf{\mu} t}$ where $\Delta_\mathbf{\mu} = \omega_\mathbf{\mu} - \omega_0$. The rapidly oscillating terms average to zero and are therefore neglected in the rotating wave approximation.

We consider transitions between the ground state, $| g_\mathbf{\mu} \rangle = | \mathcal{F} \rangle \otimes | 1_\mathbf{\mu} \rangle$, where $| 1_\mathbf{\mu} \rangle$ is a single-photon state, and the excited state, $| e \rangle = | X \rangle \otimes | 0 \rangle$, where $| X \rangle$ is given by Eq.~(\ref{eq:ExcitonExpansion}) and $| 0 \rangle$ denotes the vacuum state. We must now solve the interaction picture Schr\"odinger equation, $\frac{\mathrm{d}}{\mathrm{d}t}|\Psi(t)\rangle = -\frac{\mathrm{i}}{\hbar} H'|\Psi(t) \rangle$, assuming that the system may be in a superposition of the two eigenstates, i.e., $|\Psi(t) \rangle = c_\mathrm{e}(t) |e\rangle + \sum_\mu c_{g_\mathbf{\mu}}(t)| g_\mathbf{\mu} \rangle$ and by projecting the result onto either $\langle e |$ or $\langle g_\mu |$, we obtain the set of equations
\begin{align}
\frac{\mathrm{d}}{\mathrm{d}t} c_\mathrm{e}(t) =& \frac{\mathrm{i}q}{\hbar m_0} \sum_{\mathbf{\mu}} \sum_{\mathbf{k}_\mathrm{c},\mathbf{k}_\mathrm{v}} \frac{\epsilon_\mathbf{\mu}}{\omega_\mathbf{\mu}} e^{-\mathrm{i}\Delta_\mathbf{\mu} t}
\tilde \chi^\ast (\mathbf{k}_\mathrm{c},\mathbf{k}_\mathrm{v}) c_{g_\mathbf{\mu}}(t)\nonumber\\
&\times \hat{\mathbf{e}}_\mathbf{\mu} \cdot \int \mathrm{d}^3\mathbf{r} \psi^\ast_{\mathrm{c},\mathbf{k}_\mathrm{c}}(\mathbf{r}) A_\mathbf{\mu}(\mathbf{r}) \mathbf{p} \psi_{\mathrm{v},\mathbf{k}_\mathrm{v}}(\mathbf{r})\label{eq_WW_01}\\
\frac{\mathrm{d}}{\mathrm{d}t} c_{g_\mathbf{\mu}}(t) =& \frac{\mathrm{i}q}{\hbar m_0} \sum_{\mathbf{k}_\mathrm{c},\mathbf{k}_\mathrm{v}} \frac{\epsilon_\mathbf{\mu}}{\omega_\mathbf{\mu}} e^{\mathrm{i}\Delta_\mathbf{\mu} t}
\tilde \chi(\mathbf{k}_\mathrm{c},\mathbf{k}_\mathrm{v}) c_\mathrm{e}(t)\nonumber\\
&\times \hat{\mathbf{e}}_\mathbf{\mu} \cdot
\int \mathrm{d}^3\mathbf{r} \psi^\ast_{\mathrm{v},\mathbf{k}_\mathrm{v}}(\mathbf{r}) A^\ast_\mathbf{\mu}(\mathbf{r}) \mathbf{p} \psi_{\mathrm{c},\mathbf{k}_\mathrm{c}}(\mathbf{r}). \label{eq_WW_02}
\end{align}

Let us first turn to the spatial integrals, which include the momentum operator $\mathbf{p}=-\mathrm{i}\hbar\nabla$. We assume that $A_\mathbf{\mu}(\mathbf{r})$ and the plane-wave part of $\psi_{\mathbf{k}}(\mathbf{r})$ are slowly varying on the length scale of the lattice constant, so that these functions can be evaluated at each lattice site $\mathbf{r}_n$ and taken outside the integral:
\begin{align}
\int \mathrm{d}^3\mathbf{r} & \psi^\ast_{\mathrm{c},\mathbf{k}_\mathrm{c}}(\mathbf{r}) A_\mathbf{\mu}(\mathbf{r}) \mathbf{p} \psi_{\mathrm{v},\mathbf{k}_\mathrm{v}}(\mathbf{r})\nonumber
\\= &V_\mathrm{UC} \sum_n \big( e^{-\mathrm{i}\mathbf{k}_\mathrm{c}\cdot \mathbf{r}} A_\mathbf{\mu}(\mathbf{r}) e^{\mathrm{i}\mathbf{k}_\mathrm{v}\cdot \mathbf{r}} \big) \Big|_{\mathbf{r}=\mathbf{r}_n}\nonumber
\\ & \times \frac{1}{V_\mathrm{UC}}\int_\mathrm{UC} \mathrm{d}^3\mathbf{r}
u^\ast_{\mathrm{c},\mathbf{k}_\mathrm{c}}(\mathbf{r}) \mathbf{p} u_{\mathrm{v},\mathbf{k}_\mathrm{v}}(\mathbf{r})\nonumber\\
=&\hat{\mathbf{e}}_\mathbf{\mu} \cdot \mathbf{p}_\mathrm{cv} \int \mathrm{d}^3\mathbf{r}  e^{-\mathrm{i}\mathbf{k}_\mathrm{c}\cdot \mathbf{r}} A_\mathbf{\mu}(\mathbf{r}) e^{\mathrm{i}\mathbf{k}_\mathrm{v}\cdot \mathbf{r}}
\label{eq:MicrosopicDipoleApproximation}
\end{align}
where $\mathrm{UC}$ denotes integration over one unit cell with volume $V_\mathrm{UC}$ and we have used the orthogonality of the Bloch functions. The last equation is obtained by noting that since the Bloch functions are periodic, the integral over $\mathrm{UC}$ is the same for all $\mathbf{r}_n$ and may be evaluated separately. Also, we have assumed that the Bloch functions depend only weakly on $\mathbf{k}$, so that they can be evaluated at $\mathbf{k} =\mathbf{0}$. The sum can then be converted back to an integral and finally we have defined the Bloch matrix element as
$\mathbf{p}_\mathrm{cv} = \frac{1}{V_{\mathrm{UC}}} \int_\mathrm{UC} \mathrm{d}^3\mathbf{r} u_{\mathrm{c},\mathbf{0}}^\ast(\mathbf{r}) \mathbf{p} u_{\mathrm{v},\mathbf{0}}(\mathbf{r})$
The Bloch matrix element is a material parameter, whose magnitude evaluates to~\cite{Davies}
$\left| \mathbf{p}_\mathrm{cv} \right|^2 = \frac{m_0 E_p(x)}{2}$,
where $E_p$ is the Kane energy~\cite{Coldren_and_Corzine}, which depends on the indium mole fraction $x$ in the $\mathrm{In_xGa_{1-x}As}$ alloy.

The summations over $\mathbf{k}$-vectors can now be carried out. At this point it is advantageous to change notation to the electron-hole picture by substituting $\mathbf{k}_\mathrm{c} \to \mathbf{k}_\mathrm{e}$ and $\mathbf{k}_\mathrm{v} \to -\mathbf{k}_\mathrm{h}$. By insertion of Eq.~(\ref{eq:MicrosopicDipoleApproximation}) and interchanging the order of integration and summation in Eqs.~(\ref{eq_WW_01}) and (\ref{eq_WW_02}) the resulting equations take the form of inverse Fourier transforms and we have
\begin{align}
\begin{split}
\frac{\mathrm{d}}{\mathrm{d}t} c_\mathrm{e}(t) =& \frac{\mathrm{i}q}{\hbar m_0} \sum_{\mathbf{\mu}} \frac{\epsilon_\mathbf{\mu}}{\omega_\mathbf{\mu}} e^{-\mathrm{i}\Delta_\mathbf{\mu} t}
c_{g_\mathbf{\mu}}(t)\\& \times \hat{\mathbf{e}}_\mathbf{\mu} \cdot \mathbf{p}_\mathrm{cv} \int \mathrm{d}^3\mathbf{r} \chi^\ast (\mathbf{r}_0,\mathbf{r},\mathbf{r}) A_\mathbf{\mu}(\mathbf{r}) \label{eq:EOM1}
\end{split}
\\
\begin{split}
\frac{\mathrm{d}}{\mathrm{d}t} c_{g_\mathbf{\mu}}(t) =& \frac{\mathrm{i}q}{\hbar m_0} \frac{\epsilon_\mathbf{\mu}}{\omega_\mathbf{\mu}} e^{\mathrm{i}\Delta_\mathbf{\mu} t}
c_\mathrm{e}(t)\\&\times \hat{\mathbf{e}}_\mathbf{\mu} \cdot \mathbf{p}_\mathrm{vc} \int \mathrm{d}^3\mathbf{r} \chi(\mathbf{r}_0,\mathbf{r},\mathbf{r}) A^\ast_\mathbf{\mu}(\mathbf{r}).\label{eq:EOM2}
\end{split}
\end{align}

By integrating Eq.~(\ref{eq:EOM2}) with respect to time, inserting the result in Eq.~(\ref{eq:EOM1}), and finally rewriting the result by multiplication with a Dirac delta function in frequency and integrating over frequency we obtain
\begin{equation}
\begin{split}
\frac{\mathrm{d}}{\mathrm{d}t} c_\mathrm{e}(t) = & -\frac{\pi q^2}{2 \hbar m_0^2 \epsilon_0 }
\int_{-\infty}^{\infty} \mathrm{d}\omega \frac{\rho_{\mathrm{NL}}(\mathbf{r}_0,\omega)}{\omega}
\\&\times
\int_0^t \mathrm{d}t' e^{-\mathrm{i}\Delta_\mathbf{\mu}(t-t')} c_\mathrm{e}(t'),\label{eq_Eq001}
\end{split}
\end{equation}
where the projected non-local interaction function is defined as
\begin{align}
\begin{split}
\rho_{\mathrm{NL}}(\mathbf{r}_0,\omega) = &\left| \mathbf{p}_\mathrm{cv} \right|^2 \sum_{\mathbf{\mu}} \left| \hat{\mathbf{e}}_\mathbf{\mu} \cdot \hat{\mathbf{e}}_\mathbf{p}\right| ^2  \int \mathrm{d}^3\mathbf{r}  \chi(\mathbf{r}_0,\mathbf{r},\mathbf{r}) A^\ast_\mathbf{\mu}(\mathbf{r})
\\ &\times \int \mathrm{d}^3\mathbf{r}' \chi^\ast(\mathbf{r}_0,\mathbf{r}',\mathbf{r}') A_\mathbf{\mu}(\mathbf{r}') \delta(\omega-\omega_\mathbf{\mu}),\label{eq:NLIF_in_terms_of_A}
\end{split}
\end{align}
where $\hat{\mathbf{e}}_\mathbf{p}$ is the unit vector parallel to $\mathbf{p}_\mathrm{cv}$. Equation \ref{eq_Eq001} is a main result of this work. It is valid beyond the Markov and DA approximations and in arbitrary optical environments; it is therefore a generalization of existing theories of dipole emitters in nanophotonic structures.

If the term $\rho_{\mathrm{NL}}(\mathbf{r}_0,\omega)/\omega$ in Eq.~(\ref{eq_Eq001}) is spectrally slowly varying over the linewidth of the emitter, we may evaluate it at the emission frequency $\omega_0$ and take it outside the integral. In this Wigner-Weisskopf approximation we obtain
$\frac{\mathrm{d}}{\mathrm{d}t} c_\mathrm{e}(t) = -\frac{\pi q^2}{2 \hbar m_0^2 \epsilon_0 }\frac{\rho_{\mathrm{NL}}(\mathbf{r}_0,\omega_0)}{\omega_0} c_\mathrm{e}(t)$,
were $\int_{-\infty}^\infty \mathrm{d}\alpha e^{-i \alpha \beta} = 2 \pi \delta(\beta)$ and $\int_{0}^\infty \mathrm{d}\alpha \delta(\alpha) = \frac{1}{2}$ have been used. In the following we will write $\omega_0$ as $\omega$ for brevity. By assuming that the exciton is initially excited ($c_\mathrm{e}(0)=1$) we obtain the radiative decay of the exciton state population
$|c_\mathrm{e}(t)|^2 = e^{-\Gamma(\mathbf{r}_0,\omega) t}$,
where the radiative decay rate is defined as
\begin{equation}
\Gamma(\mathbf{r}_0,\omega) = \frac{\pi q^2}{\hbar m_0^2 \epsilon_0}
\frac{\rho_{\mathrm{NL}}(\mathbf{r}_0,\omega)}{\omega}.\label{eq_Eq003}
\end{equation}
Since $\Gamma(\mathbf{r}_0,\omega)$ depends on the exciton envelope function through the projected non-local interaction function, it is not possible to state in general whether the decay rate will increase or decrease when calculated beyond the DA; the decay rate must be calculated for a given exciton state in a given dielectric environment.
The physical significance of this result is clearer when expressed in terms of dyadic Green's tensors. This relation is derived in Appendix~\ref{sec:GreensTensors} and from Eqs.~(\ref{eq:NLIF_in_terms_of_A}) and (\ref{eq:NLIF_equivalence_of_A_and_G}) we obtain the important result
\begin{align}
\begin{split}
\rho_{\mathrm{NL}}(\mathbf{r}_0,\omega) = & \frac{2 \omega}{\pi c^2} \left| \mathbf{p}_\mathrm{cv} \right|^2 \int \mathrm{d}^3\mathbf{r} \int \mathrm{d}^3\mathbf{r}'
\chi (\mathbf{r}_0,\mathbf{r},\mathbf{r})  \chi^\ast(\mathbf{r}_0,\mathbf{r}',\mathbf{r}')
\\ &\times
 \left( \hat{\mathbf{e}}_\mathbf{p}^T \cdot \mathrm{Im}\left\{\mathbf{G}( \mathbf{r},\mathbf{r}',\omega ) \right\} \cdot \hat{\mathbf{e}}_\mathbf{p} \right).\label{eq:NLIF_in_terms_of_G}
 \end{split}
\end{align}
The Green's tensor is a propagator of the electromagnetic field. Thus, $\mathbf{G}( \mathbf{r},\mathbf{r}',\omega)$ may be interpreted as the field amplitude evaluated at the position $\mathbf{r}$ due to a dipole at $\mathbf{r}'$ with frequency $\omega$. This is illustrated in Fig.~\ref{fig:ComparisonBetweenLWAandBeyondLWA}. To actually calculate $\rho_{\mathrm{NL}}(\mathbf{r}_0,\omega)$, we must obtain the Green's tensor describing the electromagnetic environment for the particular geometry.

It is important to stress that the formalism developed here does not change the selection rules for optical transitions, i.e., they are governed by the usual dipole selection rule according to which the change in angular momentum in the transition must be $\Delta m = \pm 1$. This is fundamentally different from atomic quadrupole transitions, where $\Delta m = \pm 2$. The reason is that the quantum states of excitons consist of both an envelope and a Bloch part. Since the electromagnetic field at optical frequencies is slowly varying over a unit cell the approximation in Eq.~(\ref{eq:MicrosopicDipoleApproximation}) is very good. This unit cell DA could in principle break down for higher frequencies of the electromagnetic field and thus lead to multipole effects at the Bloch-function level but this is not relevant for the systems studied here. We note that a classical calculation of the dissipation rate of an extended dipole emitter leads to the same form of the non-local response as considered above; this is discussed in further detail in appendix~\ref{sec:ClassicalTheory}.

\section{The dipole approximation\label{sec:TheDipoleApproximation}}
Before exploring the effects beyond the DA it is instructive to consider the DA limit of the expressions derived above. In this case we can evaluate the Green's tensor at the center coordinate of the exciton, $\mathbf{r}_0$. We obtain
\begin{equation}
\begin{split}
\rho_{\mathrm{NL}}(\mathbf{r}_0,\omega) =& \frac{2 \omega}{\pi c^2} \left| \mathbf{p}_\mathrm{cv} \right|^2 \left( \hat{\mathbf{e}}_\mathbf{p}^T \cdot \mathrm{Im}\left\{\mathbf{G}( \mathbf{r}_0,\mathbf{r}_0,\omega ) \right\} \cdot \hat{\mathbf{e}}_\mathbf{p} \right)\\ &\times \left| \int \mathrm{d}^3\mathbf{r} \chi (\mathbf{r}_0,\mathbf{r},\mathbf{r}) \right|^2.\label{eq:NLIF_in_terms_of_G_LWA}
\end{split}
\end{equation}
Obviously, the integrals over the envelope functions depend only on the excitonic degrees of freedom and it is therefore natural to redefine Eq.~(\ref{eq_Eq003}) as
\begin{equation}
\Gamma_\mathrm{DA}(\mathbf{r}_0,\omega) = \frac{\pi q^2}{\hbar m_0^2 \epsilon_0} \left| \mathbf{p}_\mathrm{cv} \right|^2 \left| \int \mathrm{d}^3\mathbf{r} \chi (\mathbf{r}_0,\mathbf{r},\mathbf{r}) \right|^2 \frac{\rho(\mathbf{r}_0,\omega)}{\omega} ,\label{eq:Gamma_rad_in_terms_of_rho}
\end{equation}
where we have introduced the familiar notion of the projected LDOS~\cite{NanoOpticsBook,Dung2000,Sprik2006},
\begin{equation}
\rho(\mathbf{r}_0,\omega) = \frac{2 \omega}{\pi c^2} \left( \hat{\mathbf{e}}_\mathbf{p}^T \cdot \mathrm{Im}\left\{\mathbf{G}( \mathbf{r}_0,\mathbf{r}_0,\omega ) \right\} \cdot \hat{\mathbf{e}}_\mathbf{p} \right).\label{eq:LDOS_in_terms_of_G}
\end{equation}
The LDOS is obtained by solving Maxwell's equations and it enters the quantum optical theory of light-matter interaction as the local density of vacuum modes that spontaneous emission can occur to.

The interaction strength between an emitter and light can be characterized by the OS denoted $f(\omega)$. We define this dimensionless quantity as the ratio of the radiative decay rate in a homogeneous medium $\Gamma_\mathrm{rad,hom}(\omega)$ to the radiative decay rate $\Gamma_\mathrm{cl}(\omega)$ of a classical harmonic oscillator of elementary charge~\cite{Siegman_laserbog}, i.e.,
\begin{eqnarray}
f(\omega) = \frac{\Gamma_\mathrm{rad,hom}(\omega)}{\Gamma_\mathrm{cl}(\omega)},\label{eq:OscStrengthDef}
\end{eqnarray}
where
\begin{equation}
\Gamma_\mathrm{cl}(\omega) = \frac{n q^2 \omega^2}{6 \pi m_0 \epsilon_0 c^3}.\label{eq:GammaClDef}
\end{equation}

We can rewrite the decay rate Eq.~(\ref{eq:Gamma_rad_in_terms_of_rho}) as
\begin{equation}
\Gamma_\mathrm{DA}(\mathbf{r}_0,\omega) = \frac{\pi q^2}{2 m_0^2 \epsilon_0} f(\omega) \rho(\mathbf{r}_0,\omega),\label{eq:Gamma_rad_in_terms_of_rho_and_OS}
\end{equation}
where we have used Eqs.~(\ref{eq:OscStrengthDef}) and (\ref{eq:GammaClDef}) to obtain the OS
\begin{equation}
f(\omega) = \frac{E_p}{\hbar \omega} \left| \int \mathrm{d}^3\mathbf{r} \chi (\mathbf{r}_0,\mathbf{r},\mathbf{r}), \right|^2\label{eq:OS1}
\end{equation}
which is independent of $\mathbf{r}_0$. The usefulness of the notion of the OS in the DA is apparent from Eq.~(\ref{eq:Gamma_rad_in_terms_of_rho_and_OS}), i.e., the decay rate is given by the product of the OS and the LDOS and thus the OS quantifies the strength with which the emitter interacts with light.

The OS can be calculated readily for the exciton models discussed in appendix~\ref{sec:confinement_regimes}. In the strong confinement regime the result is
\begin{eqnarray}
f(\omega) = \frac{E_p}{\hbar \omega} \left| \int \mathrm{d}^3\mathbf{r} F_\mathrm{e}(\mathbf{r}_0,\mathbf{r}) F_\mathrm{h}(\mathbf{r}_0,\mathbf{r}) \right|^2, \label{eq:OS_DA_SC}
\end{eqnarray}
where $F_\mathrm{e}(\mathbf{r}_0,\mathbf{r})$ and $F_\mathrm{h}(\mathbf{r}_0,\mathbf{r})$ describe the independent electron and hole envelope functions, respectively. Thus, we obtain the well-known strong-confinement result~\cite{Efros1982,Andreani1999,Stobbe2010} in which the OS is proportional to the overlap of the electron and hole wave functions. As opposed to the result obtained in the single-particle picture~\cite{Stobbe2009,Stobbe2011pss}, there is no complex conjugation of either $F_\mathrm{e}(\mathbf{r}_0,\mathbf{r})$ or $F_\mathrm{h}(\mathbf{r}_0,\mathbf{r})$, which is a result of the two-particle formalism used here~\cite{Efros1982,Hanamura1988,Kayanuma1988}. The wave function overlap integral in Eq.~(\ref{eq:OS_DA_SC}) cannot exceed unity~\cite{Stobbe2010,Stobbe2011pss} and hence the maximum OS in the strong confinement regime is given by $f_\mathrm{max} = \frac{E_p}{\hbar \omega}$, which shows that $\frac{E_p}{\hbar \omega}$ can be interpreted as the OS of the bulk crystal without confinement and exciton effects.

For the spherical exciton in the weak-confinement regime,
\begin{equation}
f(\omega) = \sqrt{\pi}\frac{E_p}{\hbar \omega} \left(\frac{L}{a_0}\right)^3,\label{eq:OS_sphere_DA}
\end{equation}
where $L$ is the exciton radius and $a_0$ is the exciton Bohr radius. This is the giant-OS effect, i.e., the OS is proportional to the volume of the exciton. It is also strongly dependent on the exciton Bohr radius and therefore it shows a strong dependence on the effective masses of the carriers. For the disc-shaped exciton in the weak-confinement regime we have
\begin{align}
f(\omega) = 8\frac{E_p}{\hbar \omega}\left( \frac{L}{a_0} \right)^2.\label{eq:OS_disc_DA}
\end{align}
In this two-dimensional model the OS is proportional to the exciton area, which in the absence of inhomogeneities inside the QD~\cite{Stobbe2010} is given by the area of the QD.

In Fig.~\ref{fig:ComparisonBetweenLWAandBeyondLWA} we compare the calculated spontaneous emission rate within and beyond the DA. Fig.~\ref{fig:ComparisonBetweenLWAandBeyondLWA}(A) shows the DA result in which the decay rate is proportional to the LDOS, cf. Eq.~(\ref{eq:LDOS_in_terms_of_G}). The classical interpretation of the LDOS is that it describes self-interference, i.e., it is the field strength at the center position of the emitter, $\mathbf{r}_0$, due to the emitted light. This is given by the propagator of the field, i.e., the Green's tensor $\mathbf{G}( \mathbf{r}_0,\mathbf{r}_0,\omega )$, which is indicated as an arrow in Fig.~\ref{fig:ComparisonBetweenLWAandBeyondLWA}(A). In the QED interpretation of spontaneous emission it is stimulated by vacuum fluctuations whose density is given by the LDOS. Spontaneous emission beyond the DA is governed by the double integral appearing in Eq.~(\ref{eq:NLIF_in_terms_of_G}), i.e., it is given by the interference between all points in space weighted by the exciton envelope function as indicated in Fig.~\ref{fig:ComparisonBetweenLWAandBeyondLWA}(B).

At this point it behooves us to clarify the criterion for the validity of the DA. For a homogeneous medium the field distribution functions take the form of plane waves, i.e., $A_\mathbf{\mu}(\mathbf{r}) = \frac{e^{\mathrm{i}\mathbf{k} \cdot \mathbf{r}}}{\sqrt{\epsilon_\mathrm{r} V}}$, where $\epsilon_\mathrm{r}$ is the dielectric constant of the material and $V$ is the quantization volume. From Eq.~(\ref{eq:NLIF_in_terms_of_A}), it is clear that the DA holds when the field distribution functions are slowly varying on the scale of the variations in the envelope functions. This is equivalent to the criterion $|\mathbf{k}|\mathbf{\mathcal{L}}\ll 1$, where $\mathcal{L}$ is the characteristic length scale of the emitter. In an inhomogeneous medium the field can be expanded in terms of plane waves. This means that there is not a unique $\mathbf{k}$ for which we can evaluate this criterion. Thus, at a given frequency for which $|\mathbf{k}|\mathcal{L}\ll 1$ holds in a homogeneous medium, it will not hold in general for all $\mathbf{k}$-components of the plane-wave expansion in an inhomogeneous medium. This indicates that the use of the DA even for small QDs embedded in photonic materials, such as plasmonic nanostructures~\cite{Andersen2011} and photonic crystals, needs further justification. In general, one must simply compare the decay rate calculated in and beyond the DA to assess if it is valid.

\section{Spontaneous emission dynamics of large quantumd dots in specific inhomogeneous media\label{sec:examples}}

\begin{figure}
\includegraphics[width=\columnwidth]{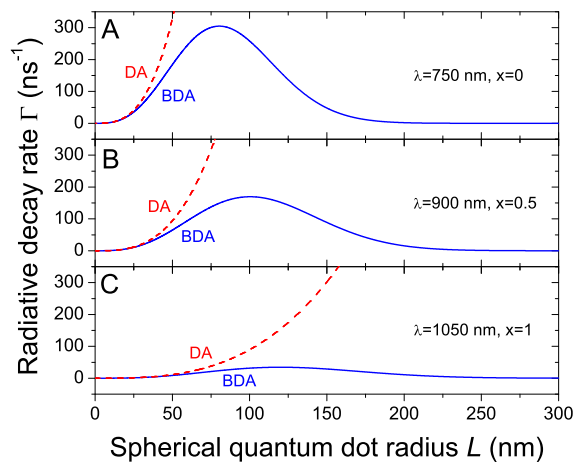}
\caption{Comparison between the radiative decay rates for spherical In$_x$Ga$_{1-x}$As QDs calculated within (dashed red lines) and beyond (solid blue lines) the DA for various emission wavelengths $\lambda$ and indium mole fractions $x$ as indicated in the figure. For large radii the decay rate is quenched.}
\label{fig:OSvsSIZEsphere}
\end{figure}

The non-local interaction function depends on a complex interplay between the specific geometry of both the electromagnetic environment and the exciton wave function and physical insight into spontaneous emission beyond the DA can be gained by considering the special cases discussed in this section. We calculate the radiative decay rate for spherical and disc-shaped excitons in homogeneous media in and beyond the DA as well as for disc-shaped excitons near a semiconductor-air interface. Beyond the DA the notion of the OS is less useful because light and matter degrees of freedom cannot be separated, i.e., Eq.~(\ref{eq:Gamma_rad_in_terms_of_rho_and_OS}) is not valid beyond the DA. We could still use Eq.~(\ref{eq:OscStrengthDef}) to obtain a dimensionless quantity characterizing the radiative decay rate in a homogeneous medium but the radiative decay rate in inhomogeneous media is neither proportional to the OS nor to the LDOS so we shall refrain from doing so.

In the strong confinement model we obtain
\begin{widetext}
\begin{equation}
\rho_{\mathrm{NL}}(\mathbf{r}_0,\omega) = \left| \mathbf{p}_\mathrm{cv} \right|^2 \frac{2 \omega}{\pi c^2} \int \mathrm{d}^3\mathbf{r} F_\mathrm{e}(\mathbf{r}_0,\mathbf{r}) F_\mathrm{h}(\mathbf{r}_0,\mathbf{r}) \int \mathrm{d}^3\mathbf{r}' F_\mathrm{e}^\ast(\mathbf{r}_0,\mathbf{r}') F_\mathrm{h}^\ast(\mathbf{r}_0,\mathbf{r}')
 \left( \hat{\mathbf{e}}_\mathbf{p}^T \cdot \mathrm{Im}\left\{\mathbf{G}( \mathbf{r},\mathbf{r}',\omega ) \right\} \cdot \hat{\mathbf{e}}_\mathbf{p} \right).\label{eq:NLIF_in_terms_of_G_strong_confinement}
\end{equation}
\end{widetext}
In systems with pronounced anisotropy between electron and hole wave functions and large optical field gradients, this can give rise to a significant orientational dependence of the radiative decay rate even for small QDs, as was recently observed experimentally~\cite{Andersen2011}. Here, however, we shall not explore this further because our focus is on large QDs.

\begin{figure}
\includegraphics[width=\columnwidth]{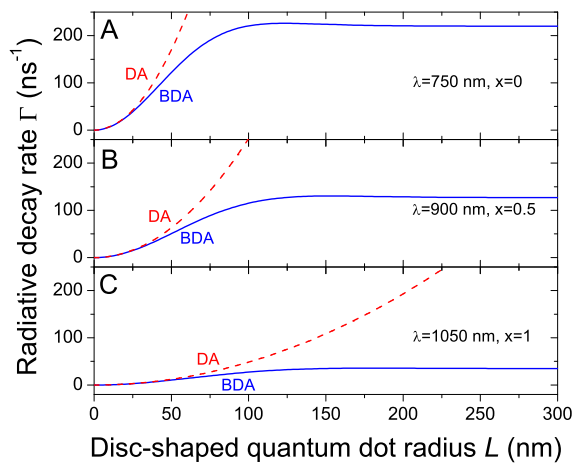}
\caption{Comparison between the radiative decay rates for disc-shaped In$_x$Ga$_{1-x}$As QDs calculated within (dashed red lines) and beyond (solid blue lines) the DA for various emission wavelengths $\lambda$ and indium mole fractions $x$ as indicated in the figure. In this case the decay rate saturates for large radii.}
\label{fig:OSvsSIZEdisc}
\end{figure}

Let us now consider spherical QDs with parabolic confinement potentials in an optically homogeneous medium. As shown in appendix~\ref{sec:AnalyticalCalculationOfTheNLIF} the non-local interaction function can be evaluated analytically in this case and the resulting decay rate is
\begin{equation}
\Gamma = \sqrt{\pi} \Gamma_\mathrm{cl} \frac{E_p}{\hbar \omega} \left(\frac{L}{a_0}\right)^3 e^{-\left(\frac{n \omega L}{2c}\right)^2},\label{eq:sphere_OS_anal}
\end{equation}
Thus, the decay rate is proportional to the bulk crystal OS, $\frac{E_p}{\hbar \omega}$, the giant-OS term, $\left(L/a_0\right)^3$, and finally $e^{-\left(\frac{n \omega L}{2c}\right)^2} = e^{-\left(\pi n L/\lambda\right)^2}$, where $\lambda$ is the vacuum wavelength of the emitted light, which is an additional term originating from the breakdown of the DA. The competition between these terms leads to a maximum in the radiative decay rate at
\begin{equation}
L_{\mathrm{max}} = \frac{\sqrt{6}\lambda}{2\pi n},\label{eq:Lmax_sphere}
\end{equation}
where
\begin{equation}
\Gamma_{\mathrm{max}}=\sqrt{6^3 \pi} \Gamma_\mathrm{cl} \frac{E_p}{\hbar \omega} \left( \frac{c}{n a_0 \omega} \right)^3.\label{eq:OSmax_sphere}
\end{equation}
Equation~(\ref{eq:sphere_OS_anal}), is plotted in Fig.~\ref{fig:OSvsSIZEsphere} along with the DA result, Eq.~(\ref{eq:OS_sphere_DA}). In a microscopically realistic model the transition energy would depend on the geometry, size, chemical composition, and strain of the QDs. Here we are not concerned with such microscopic details and we simply take the chemical composition and transition energy as being mutually independent and constant parameters. Thus, in Fig.~\ref{fig:OSvsSIZEsphere} we vary both in realistic combinations as indicated in the figure. We assume that the refractive index of the QD can be approximated by that of the surrounding medium. We describe the surrounding medium as GaAs and include the frequency dependence of the refractive index as described in Ref.~\onlinecite{Gehrsitz2000}. For simplicity we consider only heavy-hole transitions and neglect the effect of strain (the axial approximation) in which case the effective mass is isotropic, see Ref.~\onlinecite{Stobbe2009} for further details. Figure~\ref{fig:OSvsSIZEsphere} illustrates the results expressed by Eqs.~(\ref{eq:Lmax_sphere}) and (\ref{eq:OSmax_sphere}), i.e., the radiative decay rate attains a maximum when $L=L_{\mathrm{max}}$ that depends strongly on $a_0$ and thereby on the indium mole fraction. This shows firstly that pure GaAs is a more promising material for achieving a large light-matter coupling strength as compared to indium-rich alloys, due to the smaller exciton Bohr radius of GaAs excitons, and secondly that for given material parameters, a fundamental limit to the light-matter interaction strength is imposed by the breakdown of the DA. A similar size-dependence has been predicted for ZnO QDs using semiclassical approaches ~\cite{Gil2002,Fonoberov2005}. Secondly it shows that even for small QD radii the DA leads to a systematic overestimation of the light-matter interaction strength.

The fact that the radiative decay rate calculated beyond the DA vanishes for a vanishing QD radius, cf. Fig.~\ref{fig:OSvsSIZEsphere} is correct but here it arises for the wrong reasons. It is an artefact of the exciton model used here because the weak confinement approximation breaks down for small radii, i.e.,
$L\gg a_0$ is not fulfilled. For the parameters in both Fig.~\ref{fig:OSvsSIZEsphere} and Fig.~\ref{fig:OSvsSIZEdisc}, $a_0$ attains a value of 12, 16, and 29~$\mathrm{nm}$, respectively, in the subfigures (A), (B), and (C). In a more realistic confinement model but in the DA, the giant-OS effect reemerges for very small QD radii because the envelope wave functions are strongly expelled from the QD, i.e., in this regime the excitons expand~\cite{Rashba1962,Andreani1999} when the QD becomes smaller. From Fig.~\ref{fig:OSvsSIZEsphere} we can thus predict that the QD size dependence of the radiative decay rate in a more realistic confinement potential would exhibit two maxima: one due to the giant-OS effect and its quenching for large QD radii as obtained in Fig.~\ref{fig:OSvsSIZEsphere} and another giant-OS effect and its quenching at very small radii. For either vanishing or infinite QD radii the decay rate vanishes due to the breakdown of the DA.

\begin{figure}[t!]
\begin{center}
\includegraphics[width=\columnwidth]{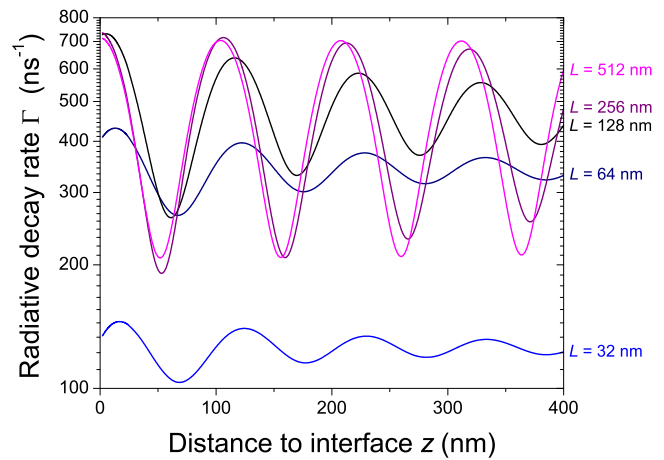}
\caption{Radiative decay rate of excitons confined in disc-shaped QDs embedded in GaAs as a function of distance to a GaAs-air interface for various QD radii $L$ as indicated in the plot. For radii up to $L=32\,$nm the distance dependence resembles the LDOS because the DA is approximately valid but for larger radii the behavior changes completely. This is a striking effect of light-matter interaction beyond the DA.
\label{fig:Oscillations}}
\end{center}
\end{figure}

For the analysis in this paper we have assumed that the subband energy level spacing, $\Delta_E$, largely exceeds the thermal energy, $k_B T$. For a relative effective heavy-hole mass of $m_\mathrm{hh}=0.59$ and $L=300\:\mathrm{nm}$ the hole subband spacing is, cf.\ Eq.~(\ref{eq:theory1:017x}), $\Delta E_\mathrm{h} = 62\:\mathrm{mK}$ so for the range of radii in Figs.~\ref{fig:OSvsSIZEsphere} and \ref{fig:OSvsSIZEdisc} the experimentally required temperatures are accessible with standard dilution refrigerators. We have also assumed that the level spacing exceeds the homogeneous linewidth of the emitter, $\hbar \Gamma_\mathrm{rad}$. This criterion is not fulfilled for all values of $L$ in Figs.~\ref{fig:OSvsSIZEsphere}, \ref{fig:OSvsSIZEdisc}, and \ref{fig:Oscillations} but could be valid in other materials in which, e.g., the transition energy would be higher. Beyond these approximations, several subbands would be populated~\cite{Takagahara1993GOSQD} and eventually the system would approach the bulk limit, which is beyond the scope of the present work.

We have numerically calculated the radiative decay rate for disc-shaped QDs and the result is shown in Fig.~\ref{fig:OSvsSIZEdisc} along with the DA result Eq.~(\ref{eq:OS_disc_DA}). Here we consider a 3\,nm thick QD, with varying lateral size. We use the same approximations as for the sphere considered above, except that here we include the anisotropy of the effective mass relevant for a strained InGaAs layer embedded in GaAs as described in Ref.~\onlinecite{Stobbe2009}. These results indicate a similar scaling of $L_{\mathrm{max}}$ and $f_{\mathrm{max}}$ as predicted by the analytical results obtained for spherical excitons. Since we keep the thickness constant, the OS does not vanish for large QD sizes as opposed to the sphere considered above. This is in agreement with results considering a non-local susceptibility of large quantum discs~\cite{Ahn2003}. Our calculation includes the numerical integrations also along the axial direction of the QD but the results are not changed significantly by assuming the DA in the $z$-direction. Also in this case the maximally achievable radiative decay rate for pure InAs QDs (Fig.~\ref{fig:OSvsSIZEdisc}(C)) is much inferior to that of GaAs QDs.

Let us now finally consider disc-shaped QDs near semiconductor-air interfaces. The interface leads to reflections, which alter the light-matter interaction. We use the same parameters as in Fig.~\ref{fig:OSvsSIZEdisc}(A) and use the Green's tensor describing the proximity of the interface~\cite{Paulus2000,Stobbe2009,NanoOpticsBook} and the result is shown in Fig.~\ref{fig:Oscillations}. For radii up to $32\,$nm the oscillations coincide with the characteristic oscillation of the LDOS~\cite{Stobbe2009} apart from a small overall reduction in the decay rate, which is consistent with Fig.~\ref{fig:OSvsSIZEdisc}(A), i.e., for small radii the DA overestimates the actual light-matter interaction strength. For $L=64\,\mathrm{nm}$ the oscillation also appears similar to the LDOS, but the DA result (not shown) is about 50\% higher than the result of the full theory carried out beyond the DA. For even larger radii ($L=128\,\mathrm{nm}$ and $256\,\mathrm{nm}$) the decay rate oscillations change dramatically. In this regime, which is far beyond the validity of the DA, the spatial dependence of the radiative decay rate exhibits pronounced deviations from the LDOS and develops into a standing wave pattern.

A comparison between Fig.~\ref{fig:OSvsSIZEdisc} and the highly non-trivial oscillations in Fig.~\ref{fig:Oscillations} leads to interesting implications for increasing the radiative decay rate. The giant-OS effect provides an effective mechanism for increasing the radiative decay rate but this effect is quenched by the breakdown of the DA. By employing optical nanostructures, the radiative decay rate can be enhanced beyond that limit. In fact, for the parameters of Fig.~\ref{fig:OSvsSIZEdisc}(A) the saturation occurs for a radiative decay rate slightly above $200\,\mathrm{ns}^{-1}$ but near a semiconductor-air interface for the same parameters (Fig.~\ref{fig:Oscillations}), the radiative decay rate can exceed $700\,\mathrm{ns}^{-1}$. This is a direct example of the intertwining of light and matter degrees of freedom imposed by the breakdown of the DA. The semiconductor-air interface considered here leads to an increase in the radiative decay rate of more than a factor of three as compared to a homogeneous medium and exploring these effects in other nanophotonic structures such as photonic crystals or optical microcavities, where the effects could be much larger, would be a very interesting future direction of research.

\section{Conclusion\label{sec:Conclusion}}

We have derived the fundamental equations governing excitonic spontaneous emission beyond the DA. The DA cannot be assumed valid \emph{a priori} in nanophotonic structures even for small QDs and thus we calculated the result for a solid state emitter beyond the DA. We derived the relation to the Green's tensor description of the electromagnetic field. In this theory the radiative decay rate of excitons is described by a non-local interaction function, which reduces to the LDOS in the dipole limit. The theory contains also the giant-OS effect in the weak confinement regime as well as the strong confinement regime as limiting cases. We have investigated and clarified the conditions under which the DA is valid. We notice that these conditions depend on both the properties of the QD, the emission wavelength, as well as the structuring of the environment and the position of the QD. Thus, the DA is in general only valid in certain points of space for a given emission energy of the QD.

Finally we note that the two cases discussed here, namely QDs embedded in either a homogenous medium or near a semiconductor-air interface, benefit from the availability of exact Green's tensors, simple experimental realization, and therefore the possibility of direct comparison between experiment and theory, but they are also the systems where the expected magnitude of the effects arising from the breakdown of the DA are smallest. A very interesting future direction would be to calculate the radiative properties of spatially extended excitons, e.g., in a photonic crystal cavity where very large differences between the DA and the theory developed here could arise.

\section{Acknowledgements}
We gratefully acknowledge financial support from The Danish Council for Independent Research (Natural Sciences and Technology and Production Sciences, projects FTP 10-080853 and FTP 10-093651), the European Research Council (ERC consolidator grant) and the Villum Kann Rasmussen Centre \emph{Natec}.

\appendix

\section{The exciton confinement regimes\label{sec:confinement_regimes}}
In this section three confinement regimes of excitons in nanostructures are discussed. The two-particle effective mass equation, Eq.~(\ref{eq:EffectiveMassEquation}), cannot in general be solved analytically for realistic QD geometries and heterostructure confinement potentials, but it can be solved within certain approximations and more importantly in different limits of the ratio between the Coulomb energy and the conduction (valence) band subband spacing, $\Delta E_\mathrm{e}$ ($\Delta E_\mathrm{h}$), in the absence of the Coulomb interaction. In the case where the Coulomb interaction is negligible, the exciton is said to be in the strong confinement regime and when it dominates over the confinement potentials, the exciton is said to be in the weak confinement regime \cite{Efros1982,Hanamura1988,Que1992,Andreani1999}.

\subsection{The unconfined regime}
In the absence of confinement, i.e., when $V_\mathrm{e}(\mathbf{r}_\mathrm{e})$ and $V_\mathrm{h}(\mathbf{r}_\mathrm{h})$ can be completely neglected, Eq.~(\ref{eq:EffectiveMassEquation}) reduces to the problem of a hydrogen atom~\cite{Bransden_and_Joachain} with effective masses. This describes a free exciton in a bulk semiconductor and in this model we can calculate the characteristic energy and length scales for an exciton. The characteristic length scale of the interparticle distance is given by the exciton Bohr radius,
\begin{eqnarray}
a_0 &=& \frac{4\pi \epsilon_0 \epsilon_\mathrm{r} \hbar^2}{q^2 m_0 m},\label{eq:theory1:018}
\end{eqnarray}
where $\epsilon_0$ denotes the vacuum permittivity, $\epsilon_\mathrm{r}$ is the relative static permittivity of the material, and the reduced mass is defined as
\begin{eqnarray}
m = \frac{m_\mathrm{e} m_\mathrm{hh}}{m_\mathrm{e} + m_\mathrm{hh}}.\label{eq:theory1:019}
\end{eqnarray}
The Coulomb potential is given by
\begin{eqnarray}
V_{\mathrm{Coul}}(\mathbf{r}_\mathrm{e},\mathbf{r}_\mathrm{h}) = - \frac{q^2}{4 \pi \epsilon_0 \epsilon_\mathrm{r} |\mathbf{r}_\mathrm{e} - \mathbf{r}_\mathrm{h}|},\label{eq:theory1:015}
\end{eqnarray}
so the energy scale, i.e., ionization energy associated with the free exciton is the effective Rydberg energy
\begin{equation}
\mathcal{R} = \frac{q^2}{4 \pi \epsilon_0 \epsilon_\mathrm{r} a_0} = \frac{\hbar^2}{m_0 m a_0^2}.\label{eq:theory1:015x}
\end{equation}

\subsection{The strong confinement regime}
In the opposite limit when Coulomb interaction may be neglected (strong confinement), the electron and hole are decoupled in Eq.~(\ref{eq:EffectiveMassEquation}), which reduces to two independent particle-in-a-box problems. These are readily solved and the solution is
\begin{equation}
\chi(\mathbf{r}_0,\mathbf{r}_\mathrm{e},\mathbf{r}_\mathrm{h}) = F_\mathrm{e}(\mathbf{r}_0,\mathbf{r}_\mathrm{e}) F_\mathrm{h}(\mathbf{r}_0,\mathbf{r}_\mathrm{h}),\label{eq:theory1:015y}
\end{equation}
where $F_\mathrm{e}(\mathbf{r}_0,\mathbf{r}_\mathrm{e})$ and $F_\mathrm{h}(\mathbf{r}_0,\mathbf{r}_\mathrm{h})$ are the electron and the hole envelope functions, respectively.
Depending on the purpose of the theoretical description, a simple model assuming infinite barriers, isotropic masses, and a simple geometry such as a cylinder or cube, may suffice. In such cases an exact solution is readily available in the literature. If we consider a cubic QD with side length $2L$, the subband spacings are~\cite{Liboff}
\begin{eqnarray}
\Delta E_{\mathrm{e}/\mathrm{h}} &=& \frac{3 \hbar^2 \pi^2}{8 m_0 m_{\mathrm{e}/\mathrm{hh}} L^2}\label{eq:theory1:017x}\\
\end{eqnarray}
In the strong confinement regime, $\mathcal{R} \ll \frac{\Delta E_\mathrm{e} + \Delta E_\mathrm{h}}{2}$, which implies that $L \ll \sqrt{\frac{3\pi^2}{16}}a_0 \approx 1.4a_0$. This simple model with infinite barriers overestimates the barrier heights and we will in general use a heuristic definition of the criteria and assume strong confinement for $L \ll a_0$ and weak confinement for $L \gg a_0$, where $2L$ is the spatial extent of the QD.

\subsection{The weak confinement regime: the spherical quantum dot}
For a spherical QD with parabolic radial confinement we have
\begin{eqnarray}
V_{\mathrm{e}/\mathrm{h}}(\mathbf{r}_0,\mathbf{r}_{\mathrm{e}/\mathrm{h}}) &=& \frac{1}{2} m_{\mathrm{e}/\mathrm{h}} \Omega^2 |\mathbf{r}_{\mathrm{e}/\mathrm{h}}-\mathbf{r}_0|^2\label{GOSQD:eq:theory1:019x}
\end{eqnarray}
where the confinement potential is given by $\Omega$ and we define the radius of the QD as $L=2\sqrt{\frac{\hbar}{M\Omega}}$. We introduce the relative and center-of-mass parameters~\cite{Que1992}
\begin{eqnarray}
\mathbf{R} &=& \frac{m_\mathrm{e} \mathbf{r}_\mathrm{e} + m_\mathrm{h} \mathbf{r}_\mathrm{h}}{m_\mathrm{e} + m_\mathrm{h}}\label{GOSQD:eq:theory1:025}\\
\mathbf{r} &=& \mathbf{r}_\mathrm{e} - \mathbf{r}_\mathrm{h}\label{GOSQD:eq:theory1:026}\\
\mathbf{P} &=& \mathbf{p}_\mathrm{e} + \mathbf{p}_\mathrm{h} \label{GOSQD:eq:theory1:027}\\
\mathbf{p} &=& \frac{m_\mathrm{h}\mathbf{p}_\mathrm{e} - m_\mathrm{e}\mathbf{p}_\mathrm{h}}{m_\mathrm{e}+m_\mathrm{h}} \label{GOSQD:eq:theory1:028}\\
M &=& m_\mathrm{e} + m_\mathrm{h}\label{GOSQD:eq:theory1:029}\\
m &=& \frac{m_\mathrm{e} m_\mathrm{h}}{m_\mathrm{e} + m_\mathrm{h}}\label{GOSQD:eq:theory1:030}.
\end{eqnarray}
By this transformation the effective-mass Hamiltonian Eq.~(\ref{eq:EffectiveMassHamiltonian}) separates into two decoupled Hamiltonians
\begin{eqnarray}
H_\mathrm{EM}(\mathbf{r}_0,\mathbf{r},\mathbf{R}) &=& H_\mathbf{R}(\mathbf{r}_0,\mathbf{R}) + H_\mathbf{r}(\mathbf{r})\label{GOSQD:eq:theory1:031}\\
H_\mathbf{R}(\mathbf{r}_0,\mathbf{R}) &=& \frac{\mathbf{P}^{2}}{2m_0M} + \frac{1}{2} M \Omega^2 |\mathbf{R}-\mathbf{r}_0|^{2}\label{GOSQD:eq:theory1:032}\\
H_\mathbf{r}(\mathbf{r}) &=& \frac{\mathbf{p}^{2}}{2m_0m} - \frac{q^2}{4 \pi \epsilon_0 \epsilon_\mathrm{r} |\mathbf{r}|}\label{GOSQD:eq:theory1:033},
\end{eqnarray}
where we have neglected the term $\frac{1}{2} m \Omega^2 |\mathbf{r}|^{2}$ in Eq.~(\ref{GOSQD:eq:theory1:033}) since we consider the weak confinement regime~\cite{Que1992}. Thus, we have reduced the problem to solving the effective mass equation for two well-known Hamiltonians, namely the three-dimensional isotropic harmonic oscillator, Eq.~(\ref{GOSQD:eq:theory1:032}), and the hydrogen problem, Eq.~(\ref{GOSQD:eq:theory1:033}). We can then write the solution to the effective mass equation as
\begin{equation}
\chi (\mathbf{r}_0,\mathbf{r}_\mathrm{e},\mathbf{r}_\mathrm{h}) = \chi' (\mathbf{r}_0,\mathbf{r},\mathbf{R}) = \chi_\mathrm{CM}(\mathbf{r}_0,\mathbf{R}) \chi_{\mathrm{rel}}(\mathbf{r})\label{eq:sphere:chi},
\end{equation}
where $\chi_\mathrm{CM}(\mathbf{r}_0,\mathbf{R})$ is the center-of-mass wavefunction and $\chi_{\mathrm{rel}}(\mathbf{r})$ is the wavefunction describing the relative motion.
For the present purposes, we are only concerned with the ground state envelope wave functions, which are given by~\cite{BransdenQuantumMechanics}
\begin{eqnarray}
\chi_\mathrm{CM}(\mathbf{r}_0,\mathbf{R}) &=& \left(\frac{2}{\pi}\right)^{3/4} \left(\frac{1}{\beta}\right)^{3/2} e^{-|\mathbf{R}-\mathbf{r}_0|^2/\beta^2}\label{eq:sphere:chi:CM}\\
\chi_{\mathrm{rel}}(\mathbf{r}) &=& \left(\frac{1}{\pi a_0^3}\right)^{1/2} e^{-|\mathbf{r}|/a_0},\label{eq:sphere:chi:rel}
\end{eqnarray}
where $\beta = \sqrt{\frac{2 \hbar}{M \Omega}}$ and $a_0$ is the exciton Bohr radius. With these definitions we have $L = \sqrt{2}\beta$ and by comparison to the definition of the normal distribution function we find that $L$ equals two standard deviations, which we define as the radius of the QD.

\subsection{The weak confinement regime: the disc-shaped quantum dot}
For a disc-shaped QD with harmonic in-plane confinement and infinite barriers in the $z$-direction we have in cylindrical coordinates, $(\rho,z,\phi)$, that~\cite{Sugawara1995}
\begin{align}
V_{\mathrm{e}/\mathrm{h}}(\mathbf{r}_0,\mathbf{r}_{\mathrm{e}/\mathrm{h}}) &=& V_{z_{\mathrm{e}/\mathrm{h}}}(z_0,z_{\mathrm{e}/\mathrm{h}}) + \frac{1}{2} m_{\mathrm{e\parallel}/\mathrm{h\parallel}} \Omega^2 |\rho_{\mathrm{e}/\mathrm{h}}-\rho_0|^2\label{eq:theory1:019x}
\end{align}
and
\begin{align}
V_{z_\mathrm{e}}(z_0,z) = V_{z_\mathrm{h}}(z_0,z) =
\Bigg\{
\begin{array}{c}
     0 \mathrm{\ for\ } |z-z_0| \leq \frac{L_z}{2}\\
\infty \mathrm{\ for\ } |z-z_0| > \frac{L_z}{2},
\end{array}
\label{eq:theory1:023}\\
\end{align}
where $L_z$ is the height of the QD. We assume that $\frac{L_z}{2} \ll a_0$ so that the Coulomb interaction in the $z$-direction may be neglected. The infinite potential in the $z$-direction is a somewhat crude approximation and in a more realistic model the wave functions would extend into the barriers. However, if we model the system with $L_z$ being slightly larger than the physical height it is a reasonable approximation although it does neglect the difference barrier penetration depths of electrons and holes due to the difference in their effective masses. This can be considered as a model of a quantum well with thickness fluctuation potentials~\cite{Gammon1996,Hours2005,Peter2005}.

For the in-plane coordinates, $\mathbf{r} = (\rho,\phi)$ and $\mathbf{p} = (p_\rho,p_\phi)$, respectively, we can make the same transformations as in Eqs.~(\ref{GOSQD:eq:theory1:025}) to (\ref{GOSQD:eq:theory1:030})~\cite{Que1992} with which the
effective-mass Hamiltonian Eq.~(\ref{eq:EffectiveMassEquation}) separates into four decoupled Hamiltonians
\begin{equation}
\begin{split}
H_\mathrm{EM}(\mathbf{r}_0,\mathbf{r},\mathbf{R},z_\mathrm{e},z_\mathrm{h})&\\ = H_\mathbf{R}(\mathbf{r}_0,\mathbf{R}) + H_\mathbf{r}&(\mathbf{r}) + H_{z_\mathrm{e}}(z_0,z_\mathrm{e}) + H_{z_\mathrm{h}}(z_o,z_\mathrm{h}),\label{eq:theory1:031}
\end{split}
\end{equation}
where
\begin{eqnarray}
H_\mathbf{R}(\mathbf{r}_0,\mathbf{R}) &=& \frac{\mathbf{P}^{2}}{2m_0M} + \frac{1}{2} M \Omega^2 |\mathbf{R}-\mathbf{r}_0|^{2}\label{eq:theory1:032}\\
H_\mathbf{r}(\mathbf{r}) &=& \frac{\mathbf{p}^{2}}{2m_0m} - \frac{q^2}{4 \pi \epsilon_0 \epsilon_\mathrm{r} |\mathbf{r}|}\label{eq:theory1:033}\\
H_{z_\mathrm{e}}(z_0,z_\mathrm{e}) &=& \frac{p_{\mathrm{e}z}^{2}}{2m_0m_{\mathrm{e}z}} + V_{z_\mathrm{e}}(z_0,z_\mathrm{e})\label{eq:theory1:034}\\
H_{z_\mathrm{h}}(z_0,z_\mathrm{h}) &=& \frac{p_{\mathrm{h}z}^{2}}{2m_0m_{\mathrm{h}z}} + V_{z_\mathrm{h}}(z_0,z_\mathrm{h})\label{eq:theory1:035},
\end{eqnarray}
where we have neglected the term $\frac{1}{2} m \Omega^2 |\mathbf{r}|^{2}$ in Eq.~(\ref{eq:theory1:033}) since we are considering the weak confinement regime~\cite{Que1992}. Thus, we have reduced the problem to solving the effective mass equation for three well-known Hamiltonians, namely those of the two-dimensional isotropic harmonic oscillator, Eq.~(\ref{eq:theory1:032}), the two-dimensional hydrogen atom, Eq.~(\ref{eq:theory1:033}), and the particle in an infinite-potential box problem, Eqs.~(\ref{eq:theory1:034}) and (\ref{eq:theory1:035}). The solution is
\begin{equation}
\begin{split}
\chi (\mathbf{r}_0,\mathbf{r},\mathbf{R},z_\mathrm{e},z_\mathrm{h})&\\
= \chi_\mathrm{CM}(\mathbf{r}_0,\mathbf{R}) &\chi_{\mathrm{rel}}(\mathbf{r}) \chi_{z_\mathrm{e}}(z_0,z_\mathrm{e}) \chi_{z_\mathrm{h}}(z_0,z_\mathrm{h}),\label{eq:DiscWF}
\end{split}
\end{equation}
where $\chi_\mathrm{CM}(\mathbf{r}_0,\mathbf{R})$ is the center-of-mass wave function, $\chi_{\mathrm{rel}}(\mathbf{r})$ is the wave function describing the relative motion, and $\chi_{z_\mathrm{e}}(z_0,z_\mathrm{e})$ and $\chi_{z_\mathrm{h}}(z_0,z_\mathrm{h})$ describe the electron and hole wave function in the $z$-direction, respectively.
The ground state envelope wave functions are given by~\cite{Liboff,Sugawara1995,Que1992}
\begin{eqnarray}
\chi_\mathrm{CM}(\mathbf{r}_0,\mathbf{R}) &=& \sqrt{\frac{2}{\pi}} \frac{1}{\beta} e^{-|\mathbf{R}-\mathbf{r}_0|^2/\beta^2}\\
\chi_{\mathrm{rel}}(\mathbf{r}) &=& \frac{4}{\sqrt{2\pi}a_0} e^{-2|\mathbf{r}|/a_0}\\
\chi_{z_\mathrm{e}}(z_0,z_\mathrm{e}) &=& \sqrt{\frac{2}{L_z}}\cos\left( \frac{\pi (z_\mathrm{e}-z_0)}{L_z} \right)\\
\chi_{z_\mathrm{h}}(z_0,z_\mathrm{h}) &=& \sqrt{\frac{2}{L_z}}\cos\left( \frac{\pi (z_\mathrm{h}-z_0)}{L_z} \right).
\end{eqnarray}

\section{Interaction between spatially extended classical emitters and classical electromagnetic fields\label{sec:ClassicalTheory}}
It is instructive to consider spontaneous emission beyond the DA in a classical model; this leads to an expression, which is very similar to the quantum result derived in sec.~\ref{sec:QuantumTheory}. A classical emitter can be described by the current density
\begin{eqnarray}
\mathbf{J}(\mathbf{r}_0,\mathbf{r}) = -\mathrm{i}\omega \bm{\mu} \rho(\mathbf{r}_0,\mathbf{r}),\label{eq_General_Source_Current}
\end{eqnarray}
where $\bm{\mu}$ is the dipole moment and $\rho(\mathbf{r}_0,\mathbf{r})$ is the density of the emitter centered at $\mathbf{r}_0$. This definition implies that the emitter is considered as a continuous distribution of infinitesimal dipoles. The power dissipation rate $\frac{\mathrm{d} W}{\mathrm{d}t}$ is given by Poynting's theorem\cite{NanoOpticsBook}
\begin{eqnarray}
\frac{\mathrm{d} W}{\mathrm{d}t} = -\frac{1}{2}\int_V \mathrm{d}^3\mathbf{r} \mathrm{Re} \big\{ \mathbf{J}^\ast(\mathbf{r}_0,\mathbf{r}) \cdot \mathbf{E}(\mathbf{r}_0,\mathbf{r}) \big\},\label{eq:Poynting}
\end{eqnarray}
where $V$ denotes the volume occupied by the emitter and $\mathbf{E}(\mathbf{r})$ is the electric field, which is given in terms of the dyadic Green's tensor $\mathbf{G}(\mathbf{r},\mathbf{r}',\omega)$ as
\begin{eqnarray}
\mathbf{E}(\mathbf{r}_0,\mathbf{r}) = \mathrm{i}\omega\mu\mu_0 \int_V \mathrm{d}^3\mathbf{r}' \mathbf{G}(\mathbf{r},\mathbf{r}',\omega)\cdot \mathbf{J}(\mathbf{r}_0,\mathbf{r}'),
\end{eqnarray}
where $\mu_0$ is the vacuum permeability. By combining these relations we readily obtain the normalized decay rate in an arbitrary dielectric environment as the ratio of power dissipation in the arbitrary structure to that in a homogeneous medium. The result is
\begin{widetext}
\begin{align}
\frac{\Gamma(\mathbf{r}_0,\omega)}{\Gamma_0(\omega)} =
\frac{\int_V \mathrm{d}^3\mathbf{r} \int_V \mathrm{d}^3\mathbf{r}' \rho(\mathbf{r}_0,\mathbf{r}) \rho(\mathbf{r}_0,\mathbf{r}') \mathbf{n}_\mu^T \cdot  \mathrm{Im} \left\{ \mathbf{G}(\mathbf{r},\mathbf{r}',\omega) \right\} \cdot  \mathbf{n}_\mu}{\int_V \mathrm{d}^3\mathbf{r} \int_V \mathrm{d}^3\mathbf{r}' \rho(\mathbf{r}_0,\mathbf{r}) \rho(\mathbf{r}_0,\mathbf{r}') \mathbf{n}_\mu^T \cdot  \mathrm{Im} \left\{ \mathbf{G}_0(\mathbf{r},\mathbf{r}',\omega) \right\} \cdot  \mathbf{n}_\mu},\label{eq_Gamma_rad_beyond_dipole_app_normalized}
\end{align}
\end{widetext}
where $\mathbf{n}_\mu$ denotes a unit vector in the direction of the polarization of the emitter and $\mathbf{G}_0(\mathbf{r},\mathbf{r}',\omega)$ denotes the Green's tensor in a homogeneous medium. The homogeneous medium decay rate, $\Gamma_0(\omega)$ does not depend on $\mathbf{r}_0$ because of the translational invariance of Greens's tensor in a homogeneous medium. In the DA we assume
\begin{eqnarray}
\rho(\mathbf{r}_0,\mathbf{r}) = \delta(\mathbf{r}-\mathbf{r}_0),
\end{eqnarray}
so Eq.~(\ref{eq_Gamma_rad_beyond_dipole_app_normalized}) reduces to the simpler and well-known result~\cite{NanoOpticsBook}
\begin{eqnarray}
\frac{\Gamma_\mathrm{DA}(\mathbf{r}_0,\omega)}{\Gamma_{0_\mathrm{DA}}(\omega)} =
\frac{\mathbf{n}_\mu^T \cdot  \mathrm{Im} \left\{ \mathbf{G}(\mathbf{r}_0,\mathbf{r}_0,\omega) \right\} \cdot \mathbf{n}_\mu}{ \mathbf{n}_\mu^T \cdot  \mathrm{Im} \left\{ \mathbf{G}_0(\mathbf{r}_0,\mathbf{r}_0,\omega) \right\} \cdot  \mathbf{n}_\mu}.
\end{eqnarray}

\section{Calculation of decay rates using Green's tensors\label{sec:GreensTensors}}
Here we derive the relation between the Green's tensor and the vector potential. The Green's tensor $\mathbf{G}(\mathbf{r},\mathbf{r}',\omega)$ for the electric field is defined as~\cite{NanoOpticsBook}
\begin{equation}
\nabla \times \nabla \times \mathbf{G}(\mathbf{r},\mathbf{r}',\omega) - \frac{\omega^2}{c^2} \epsilon_\mathrm{r}(\mathbf{r}) \mathbf{G}(\mathbf{r},\mathbf{r}',\omega) = \mathbf{I} \delta(\mathbf{r}-\mathbf{r}'),\label{eq:GreensTensorDefinition}
\end{equation}
where $\mathbf{I}$ is the identity matrix. Since the field distribution functions describe the spatial part of the solutions to Maxwell's equations, they satisfy the wave equation
\begin{equation}
\nabla \times \nabla \times \left( \mathbf{\hat{e}}_\mathbf{\mu} A_\mathbf{\mu}(\mathbf{r}) \right) -\frac{\omega_{\mathbf{\mu}}^2}{c^2} \epsilon_\mathrm{r}(\mathbf{r}) \left( \mathbf{\hat{e}}_\mathbf{\mu} A_\mathbf{\mu}(\mathbf{r}) \right) = \mathbf{0}\label{eq:WaveEquation}
\end{equation}
and the orthogonality relation
\begin{equation}
\int \mathrm{d}^3\mathbf{r} \epsilon_\mathrm{r}(\mathbf{r}) \left( \mathbf{\hat{e}}_\mathbf{\mu} A_\mathbf{\mu}(\mathbf{r}) \right) \cdot \left( \mathbf{\hat{e}}_\mathbf{\mu'} A^\ast_\mathbf{\mu'}(\mathbf{r}) \right) = \delta_{\mathbf{\mu},\mathbf{\mu}'}.
\end{equation}
Therefore we can expand the Green's tensor in terms of these functions
\begin{equation}
\mathbf{G}(\mathbf{r},\mathbf{r}',\omega) = \sum_\mathbf{\mu} \mathbf{\chi}_\mathbf{\mu}(\mathbf{r},\omega)   \left( \mathbf{\hat{e}}_\mathbf{\mu} A_\mathbf{\mu}(\mathbf{r}') \right),\label{eq:GreensFunctionExpansion}
\end{equation}
where $\mathbf{\chi}_\mathbf{\mu}(\mathbf{r},\omega)$ are expansion coefficients. By combination of Eqs.~(\ref{eq:GreensTensorDefinition}) to (\ref{eq:GreensFunctionExpansion}) we obtain
\begin{equation}
\mathbf{G}(\mathbf{r},\mathbf{r}',\omega) = \sum_\mathbf{\mu} c^2 \frac{\left( \mathbf{\hat{e}}_\mathbf{\mu} A^\ast_\mathbf{\mu}(\mathbf{r}) \right) \otimes \left( \mathbf{\hat{e}}_\mathbf{\mu} A_\mathbf{\mu}(\mathbf{r}') \right)}{\omega_\mathbf{\mu}^2-\omega^2},\label{eq:GreensTensor}
\end{equation}
where $\otimes$ denotes the dyadic product. Using the identity
\begin{align}
\begin{split}
\lim_{\eta \to 0} \mathrm{Im} \left\{ \frac{1}{\omega_\mathbf{\mu}^2-(\omega+\mathrm{i}\eta)^2} \right\} = &\\ \frac{\pi}{2\omega_\mathbf{\mu}} (\delta(\omega-\omega_\mathbf{\mu}) & -\delta(\omega + \omega_\mathbf{\mu}) ),
\end{split}
\end{align}
multiplying with $\left( \mathbf{\hat{e}}_\mathbf{\mu} A^\ast_\mathbf{\mu}(\mathbf{r}) \right) \otimes \left( \mathbf{\hat{e}}_\mathbf{\mu} A_\mathbf{\mu}(\mathbf{r}') \right)$, and summing over all $\mathbf{\mu}$ we obtain the useful relation
\begin{align}
\begin{split}
\lim_{\eta \to 0} \mathrm{Im} \left\{ \sum_\mathbf{\mu} \frac{\left( \mathbf{\hat{e}}_\mathbf{\mu} A^\ast_\mathbf{\mu}(\mathbf{r}) \right) \otimes \left( \mathbf{\hat{e}}_\mathbf{\mu} A_\mathbf{\mu}(\mathbf{r}') \right)}{\omega_\mathbf{\mu}^2-(\omega+\mathrm{i}\eta)^2}  \right\} &\\
= \frac{\pi}{2\omega} \sum_\mathbf{\mu} \left( \mathbf{\hat{e}}_\mathbf{\mu} A^\ast_\mathbf{\mu}(\mathbf{r}) \right) \otimes \left( \mathbf{\hat{e}}_\mathbf{\mu} A_\mathbf{\mu}(\mathbf{r}') \right) & \delta(\omega-\omega_\mathbf{\mu}).\label{eq:GreensTensorUsefulRelation}
\end{split}
\end{align}
Here we have discarded the unphysical delta function $\delta(\omega+\omega_\mathbf{\mu})$. Now, from Eqs.~(\ref{eq:GreensTensor}) and (\ref{eq:GreensTensorUsefulRelation}) we obtain
\begin{align}
\begin{split}
\mathrm{Im} \left\{ \mathbf{G}(\mathbf{r},\mathbf{r}',\omega) \right\} = &\\
\frac{\pi c^2}{2 \omega}
\sum_\mathbf{\mu} \big( \mathbf{\hat{e}}_\mathbf{\mu} & A^\ast_\mathbf{\mu}(\mathbf{r}) \big) \otimes \left( \mathbf{\hat{e}}_\mathbf{\mu} A_\mathbf{\mu}(\mathbf{r}') \right) \delta(\omega-\omega_\mathbf{\mu}).\label{eq:ImGInTermsOfA}
\end{split}
\end{align}
From this result the equivalence of Eq.~(\ref{eq:NLIF_in_terms_of_A}) and Eq.~(\ref{eq:NLIF_in_terms_of_G}) can be found directly by performing a series of operations on both sides of the equation. By projecting onto $\hat{\mathbf{e}}_\mathbf{p}$ from left and right, multiplication with $\frac{2\omega}{\pi c^2} |\mathbf{p}_\mathrm{cv}|^2 \chi(\mathbf{r}_0,\mathbf{r},\mathbf{r}) \chi^\ast(\mathbf{r}_0,\mathbf{r}',\mathbf{r}')$, and finally integration over both $\mathbf{r}$ and $\mathbf{r}'$ we obtain
\begin{widetext}
\begin{align}
\begin{split}
\frac{2 \omega}{\pi c^2} \left| \mathbf{p}_\mathrm{cv} \right|^2 \int \mathrm{d}^3\mathbf{r} \int \mathrm{d}^3\mathbf{r}' & \chi (\mathbf{r}_0,\mathbf{r},\mathbf{r}) \chi^\ast(\mathbf{r}_0,\mathbf{r}',\mathbf{r}') \left( \hat{\mathbf{e}}_\mathbf{p}^T \cdot \mathrm{Im}\left\{\mathbf{G}( \mathbf{r},\mathbf{r}',\omega ) \right\} \cdot \hat{\mathbf{e}}_\mathbf{p} \right)
=\\
&\left| \mathbf{p}_\mathrm{cv} \right|^2 \sum_{\mathbf{\mu}} \left| \hat{\mathbf{e}}_\mathbf{\mu} \cdot \hat{\mathbf{e}}_\mathbf{p}\right| ^2  \int \mathrm{d}^3\mathbf{r}  \chi(\mathbf{r}_0,\mathbf{r},\mathbf{r}) A^\ast_\mathbf{\mu}(\mathbf{r}) \int \mathrm{d}^3\mathbf{r}' \chi^\ast(\mathbf{r}_0,\mathbf{r}',\mathbf{r}') A_\mathbf{\mu}(\mathbf{r}') \delta(\omega-\omega_\mathbf{\mu})
\label{eq:NLIF_equivalence_of_A_and_G}.
\end{split}
\end{align}
\end{widetext}
Thus, the right-hand sides of Eq.~(\ref{eq:NLIF_in_terms_of_A}) and Eq.~(\ref{eq:NLIF_in_terms_of_G}) are identical.

\section{Explicit evaluation of matrix elements for gaussian wave functions\label{sec:AnalyticalCalculationOfTheNLIF}}
Here we show the explicit analytical evaluation of the non-local interaction function Eq.~(\ref{eq:NLIF_in_terms_of_G}) for spherical excitons in the weak confinement regime. By insertion of Eqs.~(\ref{eq:sphere:chi}), (\ref{eq:sphere:chi:CM}), and (\ref{eq:sphere:chi:rel}) in Eq.~(\ref{eq:NLIF_in_terms_of_G}) we see that the integral to be solved is of the form
\begin{align}
I^{\alpha\alpha} = \int \mathrm{d}^3\mathbf{r}\int \mathrm{d}^3\mathbf{r}' f(\mathbf{r})f^*(\mathbf{r}')\mathrm{Im}\left\{\mathbf{e}_\alpha^T \cdot\mathbf{G}(\mathbf{r},\mathbf{r}')\cdot\mathbf{e}_\alpha \right\},
\label{Eq:nonLocalInteraction_GeneralForm}
\end{align}
where $f(\mathbf{r}) = f_0e^{-|\mathbf{r}|^2/\beta^2}$ and $f_0^2= \sqrt{8}\pi^{-5/2}a_0^{-1}\beta^{-3}$. For a homogeneous medium, no generality is lost by choosing $\mathbf{r}_0=\mathbf{0}$ and we have therefore suppressed $\mathbf{r}_0$ in the following. This allows for an explicit evaluation of the non-local decay function for homogeneous media. In the case of non-homogeneous media, we can always express the Green's tensor, and hence the integral in Eq.~(\ref{Eq:nonLocalInteraction_GeneralForm}), as the sum of a homogeneous part and a scattering part so this calculation is useful also for inhomogeneous media.

Using the expression for the Green's tensor in homogeneous media \cite{Martin1998} we can rewrite Eq.~(\ref{Eq:nonLocalInteraction_GeneralForm}) as
\begin{align}
\begin{split}
I^{\alpha\alpha}
=&f_0^2 \int\mathrm{d}^3\mathbf{r} \int\mathrm{d}^3\mathbf{r}'
e^{-r^2/\beta^2}e^{-r'^2/\beta^2}\\
&\times\mathrm{Im}\left\{\left(\mathbf{I}+\frac{1}{k^2} \frac{\partial^2}{\partial \alpha^2} \right) \frac{k}{4\pi}h_0(k\xi)\right\}
\end{split}\\
\begin{split}
=&f_0^2 \int\mathrm{d}^3\mathbf{r} \int\mathrm{d}^3\mathbf{r}'
e^{-r^2/\beta^2}e^{-r'^2/\beta^2}\\
&\times\left(\delta_{\alpha,\alpha}+\frac{1}{k^2} \frac{\partial^2}{\partial \alpha^2} \right) \frac{k}{4\pi}j_0(k\xi)
\end{split}
\end{align}
in which $\xi=|\mathbf{r}-\mathbf{r}'|$, $r=|\mathbf{r}|$, and $k=|\mathbf{k}|$ is the magnitude of the wave vector in the background material, and $j_0$ and $h_0$ denote the spherical Bessel and Hankel functions of the first kind, respectively. The spherical Bessel function can be rewritten in terms of other spherical Bessel functions and spherical harmonics as~\cite{Martin2006}
\begin{align}
\begin{split}
j_0(k\xi) &=\\  4\pi\sum_{n=0}^\infty&\sum_{m=-n}^n (-1)^m j_n(k r')Y_n^{-m}(\theta',\phi')j_n(k r)Y_n^m(\theta,\phi)
\end{split}
\end{align}
in which the spherical harmonics are defined as
\begin{align}
Y_n^m(\theta,\phi) = (-1)^m\sqrt{\frac{(2n-1)(n-m)!}{4\pi(n+m)!}}P_n^m(\cos\theta)e^{\mathrm{i}m\varphi},
\end{align}
where $P_n^m(\cos\theta)$ is the associated Legendre function. We now perform the angular integration over $\mathbf{r}'$ to find $m=0$ and $n=0$. In addition, we note that $\frac{\partial^2}{\partial \alpha^2}$ acts only on $j_0(k r)$. Following Ref.~\onlinecite{Martin2006} we now write $j_n(k r)Y_n^m(\mathbf{r})=\Omega_n^m$ in which case we may express the derivative in terms of raising and lowering operators $D_+$ and $D_-$, respectively, defined as
\begin{align}
D_\pm=-\frac{1}{k}\left(\frac{\partial}{\partial x}\pm \mathrm{i}\frac{\partial}{\partial y}\right),
\end{align}
and with the actions (for $0\leq|m|\leq n$):
\begin{subequations}
\begin{align}
D_+\Omega^m_n &= -\sqrt{\frac{(n+m+2)(n+m+1)}{(2n+1)(2n+3)}}\Omega^{m+1}_{n+1} \nonumber \\
&\quad -\sqrt{\frac{(n-m)(n-m-1)}{4n^2-1}}\Omega^{m+1}_{n-1} \\
D_-\Omega^m_n &= \sqrt{\frac{(n-m+2)(n-m+1)}{(2n+1)(2n+3)}}\Omega^{m-1}_{n+1} \nonumber \\
&\quad +\sqrt{\frac{(n+m)(n+m-1)}{4n^2-1}}\Omega^{m-1}_{n-1}\\
\end{align}
\end{subequations}
For the $\alpha = x$ term we find
\begin{align}
\frac{1}{k^2}\frac{\partial^2}{\partial x^2} = \frac{1}{4}\left(D_+^2 + D_-^2 + 2D_+D_- \right).
\end{align}
Only the $D_+D_-$ term results in non-vanishing terms after angular integration and we have
\begin{align}
D_+\Omega_0^0 &= -\sqrt{\frac{2}{3}}\Omega_1^1 \\
-\sqrt{\frac{2}{3}}D_-\Omega_1^1 &= -\frac{2}{3}\Omega_0^0 - \frac{2}{\sqrt{45}}\Omega_2^0.
\end{align}
The angular integral over $\Omega_2^0$ vanishes, leaving only the $\Omega_0^0$ term. In this way we obtain the final expression for $I^{xx}$ as
\begin{align}
I^{xx}&=\frac{1}{6} f_0^2 k^2\pi^2\beta^6e^{-k^2\beta^2/2}.
\end{align}

\end{document}